\documentclass[preprint,prd,showpacs,nofootinbib]{revtex4}

\usepackage{graphicx}
\usepackage{amsmath,amsfonts,amssymb}
\def\negcdot{\negmedspace\cdot\negmedspace}

\newcommand*{\chpt}{\raise0.4ex\hbox{$\chi$}PT}
\newcommand*{\schpt}{S\raise0.4ex\hbox{$\chi$}PT}
\newcommand*{\ie}{\textit{i.e.},\ }
\newcommand*{\eg}{\textit{e.g.},\ }
\newcommand*{\etc}{\textit{etc.}}
\newcommand*{\via}{\textit{via}}
\newcommand*{\et}{\textit{et al.}}
\def\cf{{\it c.f.}\ }

\providecommand*{\prd}[1]{Phys.\ Rev.\ \textbf{D#1}}
\renewcommand*{\prd}[1]{Phys.\ Rev.\ \textbf{D#1}}

\newcommand*{\npbps}[1]{Nucl.\ Phys.\ \textbf{B}
        (Proc.\ Suppl.) \textbf{#1}}

\newcommand*{\berlin}{\npbps{106} (2002)}
\newcommand*{\boston}{\npbps{119} (2003)}
\newcommand*{\tsukuba}{\npbps{129-130C} (2004)}
\def\fermilabtwo{Nucl.\ Phys.\ {\bf B} (Proc.\ Suppl.) {\bf 140} (2005)}

\newcommand*{\MeV}{{\rm Me\!V}}
\newcommand*{\GeV}{{\rm Ge\!V}}
\newcommand*{\fm}{{\rm fm}}

\newcommand{\opopo}{\ensuremath{1\!+\!1\!+\!1}}
\newcommand{\fpfpf}{\ensuremath{4\!+\!4\!+\!4}}

\newcommand*{\Tr}{\ensuremath{\operatorname{Tr}}}

\newcommand{\trD}{\ensuremath{\textrm{tr}_{\textrm{\tiny \it D}}}}

\def\gtwid{{\,\raise.35ex\hbox{$>$\kern-.75em\lower1ex\hbox{$\sim$}}\,}}
\def\ltwid{{\,\raise.35ex\hbox{$<$\kern-.75em\lower1ex\hbox{$\sim$}}\,}}
\def\leftvec{{\raise1.5ex\hbox{$\leftarrow$}\kern-1.00em}}
\def\rightvec{{\raise1.5ex\hbox{$\rightarrow$}\kern-1.00em}}
\def\half{{\scriptstyle \raise.2ex\hbox{${1\over2}$}}}
\def\threehalves{{\scriptstyle \raise.15ex\hbox{${3\over2}$}}}
\def\third{{\scriptstyle \raise.15ex\hbox{${1\over3}$}}}
\def\third{{\scriptstyle \raise.15ex\hbox{${1\over3}$}}}
\def\twothirds{{\scriptstyle \raise.15ex\hbox{${2\over3}$}}}
\def\fourth{{\scriptstyle \raise.15ex\hbox{${1\over4}$}}}

\newcommand{\Dslash}{\ensuremath{D\!\!\!\! /}}

\newcommand{\vslash}{\ensuremath{v\!\!\! /}}

\newcommand{\cD}{\ensuremath{\mathcal{D}}}

\newcommand{\cI}{\ensuremath{\mathcal{I}}}

\newcommand{\cL}{\ensuremath{\mathcal{L}}}
\newcommand{\cM}{\ensuremath{\mathcal{M}}}

\newcommand{\cO}{\ensuremath{\mathcal{O}}}

\newcommand{\cT}{\ensuremath{\mathcal{T}}}

\newcommand{\cV}{\ensuremath{\mathcal{V}}}

\newcommand{\eq}[1]{Eq.~\eqref{eq:#1}}
\newcommand{\Eq}[1]{Equation \eqref{eq:#1}}
\newcommand{\eqs}[2]{Eqs.~\eqref{eq:#1} and \eqref{eq:#2}}

\newcommand{\eqsthru}[2] {Eqs.~\eqref{eq:#1} through \eqref{eq:#2}}
\newcommand{\eqsor}[2]{Eq.~\eqref{eq:#1} or \eqref{eq:#2}}
\newcommand{\eqsthree}[3]{Eqs.~\eqref{eq:#1}, \eqref{eq:#2} and \eqref{eq:#3}}
\newcommand{\Eqsthree}[3]{Equations~\eqref{eq:#1}, \eqref{eq:#2}, and \eqref{eq:#3}}

\def\figref#1{Fig.~\ref{fig:#1}}

\def\secref#1{Sec.~\ref{sec:#1}}
\def\secrefs#1#2{Secs.~\ref{sec:#1} and \ref{sec:#2}}
\def\secthru#1#2{Secs.~\ref{sec:#1} to \ref{sec:#2}}

\begin{document}

\title{Staggered Chiral Perturbation Theory for Heavy-Light Mesons}
\author{C.\ Aubin}
\affiliation{Dept.\ of Physics,
Columbia University, New York, NY 10027}
\author{C.\ Bernard}
\affiliation{Washington University, St.\ Louis, MO 63130}
\begin{abstract}
We incorporate heavy-light mesons into staggered chiral perturbation theory (\schpt),
working to leading order in $1/m_Q$, where $m_Q$ is the heavy quark mass.
At first non-trivial order in the chiral expansion, staggered taste violations affect the
chiral logarithms for heavy-light quantities only through the light meson propagators in loops.
There are also new analytic contributions coming
from additional terms in the Lagrangian
involving heavy-light and light mesons.
Using this heavy-light \schpt,
we perform the one-loop calculation of the $B$ (or $D$) meson leptonic decay 
constant in the partially quenched and full QCD cases.
In our treatment, we assume the validity both of the ``fourth root trick'' 
to reduce four staggered tastes to one, and of the \schpt\ prescription to represent
this trick by insertions of factors of $1/4$ for each sea quark loop.
\end{abstract}
\pacs{12.39.Fe,12.39.Hg, 11.30.Rd, 12.38.Gc}

\maketitle

\section{Introduction}\label{sec:intro}

The lattice can make a major contribution to the understanding of
flavor physics through the computation of the properties of
heavy-light mesons. 
Among the important quantities to calculate are the heavy-light decay constants,
form factors, and $B$ parameters. (See Ref.~\cite{WINGATE-REVIEW,KRONFELD-REVIEW} for recent reviews.) 
However, the ``traditional'' lattice approach,
using quite massive light quarks, is likely to produce results that
have large systematic errors coming from the long
chiral extrapolation \cite{KRONFELDandRYAN}. An alternative
approach that seems very promising is to use staggered (Kogut-Susskind)
fermions for both the light valence and sea 
quarks \cite{WINGATE,WINGATE_2,WINGATE_3,SHIGEMITSU,Aubin:2004ej,Aubin:2005ar}.
An exact chiral symmetry for staggered quarks at finite lattice spacing  allows
simulations to be performed at small quark mass, deep in the chiral regime.

Staggered quarks have ``taste'' symmetry, a four-fold remnant of the doubling symmetry.
Taste symmetry is broken at $\cO(a^2)$ in the lattice
spacing, resulting in rather large discretization effects. 
Even at the smaller lattice spacing employed in current MILC
simulations with improved staggered quarks \cite{FPI04} (the ``fine'' set with lattice spacing $a\approx 0.09\ \fm$), the effects are not negligible in
any quantity which is sensitive to pseudoscalar meson loops.
This has been understood in the light
meson sector using staggered chiral perturbation theory (\schpt)
\cite{LEE_SHARPE,CHIRAL_FSB,SCHPT}. We now extend this program to include heavy
quarks, thus merging heavy quark
effective theory (HQET) with \schpt. We then calculate the leptonic decay constants, $f_B$ and $f_{B_s}$,
at one loop in \schpt\ in both the partially quenched and full QCD cases.
This is one of the simpler quantities to calculate both in
lattice simulations \cite{WINGATE-REVIEW,KRONFELD-REVIEW} and in the \schpt\ formalism. 

Simulations with staggered fermions such as
Refs.~\cite{WINGATE,WINGATE_2,WINGATE_3,SHIGEMITSU,Aubin:2004ej,Aubin:2005ar,FPI04},
use the  ``fourth root trick'' \cite{Marinari:1981qf},
designed to reduce the number of tastes per flavor from 4 to 1 in the continuum limit. 
The validity of the fourth root trick has not been proven, although 
various recent studies  have, in our opinion, made it rather plausible
\cite{FOURTH-ROOT-TEST}.
In the following, we {\it assume} that this trick is valid, and that it can be represented
in the chiral theory by insertions of factors of $1/4$ for each sea quark 
loop \cite{SCHPT,CHIRAL_FSB}. We employ 
a quark-flow analysis \cite{Sharpe:1992ft} to locate these sea quark loops.
The ``replica method'' \cite{Damgaard:2000gh} is an 
alternative technique for inserting the appropriate factors.
It is theoretically cleaner than quark flow, but usually somewhat more complicated to
implement in practice.  One would start
with $n_S$ staggered fields for each continuum sea-quark flavor desired, and take the limit
$n_S\to 1/4$ at the end. The two approaches yield identical results in the current
case, and in all other cases that have been investigated to date.  

A brief discussion of the current results, as well
as the application to semileptonic form factors, was previously presented
in Ref.~\cite{Aubin:2004xd}. A detailed description of the
semileptonic calculation is in preparation \cite{CA-CB-SEMILEPTONIC}. 
The \schpt\ forms obtained have already been used for the
chiral (and continuum) extrapolations of numerical lattice results for heavy-light 
semileptonic \cite{Aubin:2004ej} and leptonic \cite{Aubin:2005ar} decays.

In this paper the heavy quark
mass $m_Q$ is taken to be large  compared to
$\Lambda_{\rm QCD}$, so that working to leading order in the heavy quark expansion
(\ie neglecting $1/m_Q$ terms) 
is a reasonable first approximation.  
We also do not include, in our effective theory, discretization errors
coming from the heavy quark.  We assume that such errors can be
adequately estimated independently, using HQET as the 
effective theory description of the lattice heavy 
quark \cite{Kronfeld:2000ck,KRONFELD-REVIEW,OKAMOTO-05}. However, in order to separate
light and heavy quark discretization errors, our analysis requires
that the heavy quark mass not be much greater than one in lattice units.
The rather counterintuitive restriction arises because the heavy
quark action no longer sufficiently suppresses doublers when $am_Q\gg 1$.
Indeed, in the large $am_Q$ limit, the heavy quark action becomes
a static lattice action, where there is no suppression of spatial doublers
at all, since a static quark's energy is independent of its momentum.
When heavy quark doublers are insufficiently suppressed,
the light and heavy quarks in a heavy-light meson can exchange gluons
of momenta $\pi/a$, leading to 
``mixed'' 4-quark operators (products of heavy and light bilinears) in the Symanzik
action
that violate taste.  
In order to avoid treating this complication --- which would inextricably mix 
light- and heavy-quark discretization errors ---
we do not allow $am_Q\gg 1$ in our analysis. On the other hand,
typical values of $am_Q$ in practical calculations with 
the Fermilab \cite{El-Khadra:1996mp} or NRQCD 
\cite{NRQCD} heavy quarks are likely to be acceptable.  With this restriction,
all taste-violations in
the $\cO(a^2)$ Symanzik action occur solely in the light quark sector; the taste-violating
terms are thus the same
as in Refs.~\cite{LEE_SHARPE,SCHPT}.
More details, including
a discussion based on the symmetries of the lattice action, appear in \secref{Symanzik}.

Incorporating heavy quarks into \schpt\ produces
a large number of terms in the chiral Lagrangian involving 
combinations of the heavy-light and the light mesons.
All of these terms,
which arise due to the taste symmetry breaking,
are however non-leading: They do not appear in one-loop 
diagrams for the heavy-light quantities and
enter the decay constant result through a single analytic term.
Non-trivial taste violations arise only in the light meson
propagators in the one-loop diagrams.  Since the 
taste-violating low energy constants 
in the light meson sector have been determined from simulations and
\schpt\ fits \cite{FPI04,LAT02-03POSTERS,MILC_SPECTRUM}, the new analytic term is the only additional
fit parameter beyond those low-energy constants that appear in the continuum,
such as the $B$-$B^*$-$\pi$ coupling $g_\pi$.

This paper is organized as follows. We first describe the Symanzik
action to $\cO(a^2)$ in \secref{Symanzik}.  The \schpt\
Lagrangian, here including heavy-light mesons, is then constructed from a
spurion analysis in
\secrefs{ls-lag}{4quark-ops}, where the latter focuses on the
$\cO(a^2)$ terms involving heavy-light mesons.
In Sec.~\ref{sec:fB} we calculate the one-loop expression for the chiral
logarithms that arise in the leptonic decay constant, $f_{B_x}$, where
$x$ is the flavor of the light valence quark. We
write down the final results for various cases in
Sec.~\ref{sec:final_results} and discuss some features of the
low energy constants (LECs) in \secref{scale}. Finite volume effects
are treated in Sec.~\ref{sec:fin_vol}. We finish with some remarks and conclusions 
in Sec.~\ref{sec:conc}.  

\section{Symanzik Action}
\label{sec:Symanzik}

For concreteness, we consider here a heavy quark that is simulated either with NRQCD \cite{NRQCD} or
with the Fermilab interpretation \cite{El-Khadra:1996mp} of a clover or more
highly improved quark.\footnote{The formalism will also apply when the heavy quark
is simulated by extrapolation up in mass from a lighter, conventional, lattice quark.}
What is crucial for us about these possible heavy quark actions is that they describe
a single physical fermion --- doubler\footnote{For convenience, we use the term
``doubler'' to describe a heavy-quark state with one or more components of 3-momentum near
$\pi/a$.  Its ``rest mass'' is defined to be the energy of the state when all 3-momentum components
are exactly $\pi/a$ or 0.  Since these states are always suppressed in the cases considered here,
we do not need to be concerned that the interpretation of such states as
additional heavy {\it particles}\/ is problematic:  With Wilson-like or non-relativistic fermions,
their energy {\it decreases}\/ when one of the large components of momentum changes away  from $\pi/a$.}
 masses are assumed to be larger than the physical
heavy quark mass by an amount of order of the cutoff, \ie $\sim 1/a$.  This means that the heavy
quark fields in the Symanzik action will have no degree of freedom corresponding to ``taste.''

In the limit $am_Q\gg1$, however, it will not be possible to neglect
the doublers with either the Fermilab or the NQRCD actions.  That is because these actions
approach a static lattice theory,
which has no intrinsic suppression of spatial doublers: The energy of
a static quark is independent of its three-momentum.  This would not be a problem if the
light quark in a heavy-light bound state were simulated as a Wilson-like fermion. 
In that case, the light-quark Wilson term would suppress gluon
exchanges of momentum $\pi/a$ between the light and heavy quarks, by forcing the 
light quark far off shell.  But emission or absorption of such a gluon is not 
suppressed by the staggered light action, and indeed is
simply a taste-changing interaction.  
To keep the analysis manageable, we
must assume that the {\em heavy}\/ quark action suppresses these exchanges.
We define $\Delta_M$ to be the splitting between (the rest masses of)
the physical heavy quark
and the closest doublers. To be able to ignore the effect of doubler states in 
the low energy effective theory, we must have
$\Delta_M \gg \Lambda_{QCD}$.
However, for the Fermilab or NRQCD action, 
$\Delta_M$ goes like $1/(a^2m_Q)$ for $am_Q\gg 1$; eventually the condition
$\Delta_M \gg \Lambda_{QCD}$ will be violated as $m_Q$ grows at fixed $a$.

We wish to correct a possible misunderstanding here. When
$a m_Q \sim 1$, one might worry that doublers cannot be neglected
relative to the physical heavy quark since both physical and
doubler masses are formally $\cO(1/a)$. However, what is important
is not (say) the ratio of doubler mass to physical mass, but the {\it splitting}\/
$\Delta_M$ between doubler masses and the physical mass. As long as
$\Delta_M \gg \Lambda_{QCD}$, the doubler states can treated as ``integrated out''
and their effects can be summarized  by higher-dimension operators in the
Symanzik/HQET theory \cite{Kronfeld:2000ck,KRONFELD-REVIEW} involving only the 
light quarks and the {\it physical} heavy quark.  Systematic errors coming
from doublers are thus included in estimates of heavy-quark discretization
effects from higher operators. (See further discussion below).

It is instructive to estimate $\Delta_M$ in some
practical situations.  For Wilson-type quarks, the rest (or ``pole'') mass $m_1$ of the physical
quark is given at tree level by $a m_1 = \ln(1+am_0)$, where $am_0=1/(2\kappa)-1/(2\kappa_{crit})$,
with $\kappa$ the hopping parameter and $\kappa_{crit}$ its critical value.  For the lowest
doubler, with momentum $\pi/a$ in a single lattice direction, the Wilson term $1-\cos(ap)$
becomes 2 instead of 0, and the doubler rest mass $m^D_1$ is given by
$a m^D_1 = \ln(3+am_0)$. For the MILC ``coarse'' lattices \cite{FPI04}, $a^{-1}\approx 1.59\; \GeV$,
$\kappa_{crit}\approx 0.1378$, and $\kappa_b$, the hopping parameter of a $b$ quark is
$\approx 0.086$ \cite{OKAMOTO}.  This gives $\Delta_M\equiv m^D_1-m_1\approx 775\; \MeV$. Alternatively,
using the tadpole improvement \cite{Lepage:1992xa} parameter $u_0\equiv 1/(8\kappa_{crit})$ gives 
$am_0 = 4(\kappa_{crit}/\kappa -1)$ and  $\Delta_M\approx 735\; \MeV$.  These values of  $\Delta_M$,
which are comparable to other masses dropped in \chpt\ (\eg $m_\rho$), are probably large enough
to neglect the effect of doublers in the current chiral analysis.  However, it is clear that applying this
analysis to any lattices that are significantly coarser than the MILC coarse lattices will be problematic.
For example, at $a^{-1}=1\; \GeV$, we estimate $\Delta_M\approx 350\; \MeV$, which is certainly
not much larger than $\Lambda_{QCD}$.  With $\Delta_M$ this low, any effective
chiral theory becomes very complicated, and there also would be practical
problems  in separating the physical state from the doubler states in simulations.

A similar analysis applies in NRQCD.  Using a bare mass $aM=2.8$ for $b$ quarks on the MILC
coarse lattices and a stabilization parameter $n=2$ \cite{WINGATE_3}, we find 
$\Delta_M\approx 930\; \MeV$, which should be adequate.  
However, with $a^{-1}=1\; \GeV$, $\Delta_M\approx 400\; \MeV$, again much too low to
omit the effects of doublers from the effective chiral theory.

If heavy quark doubler effects may be neglected,
the analysis of the Symanzik action is very similar
to that in Ref.~\cite{BBRS}, which considers 
Ginsparg-Wilson valence quarks coupled to staggered sea quarks.
Taste violations at $\cO(a^2)$ appear only in four-quark operators composed
exclusively of light (staggered) quarks.  These terms in the Symanzik
action are identical to those in \cite{LEE_SHARPE,SCHPT}. 
``Mixed'' four-quark operators consisting of the product of a light quark bilinear and a heavy
quark bilinear do not break taste symmetry. Physically, operators that violate taste symmetry
require momentum $\pi/a$ gluon exchanges, which we ban from the Symanzik action when we
omit doubler states from the low-energy theory. A more rigorous proof that follows from the
symmetries of the lattice action can also be constructed. 
 The proof is identical to that presented 
in Ref.~\cite{BBRS}, so we do not include it in detail here.  
The basic idea is that the continuum, physical, heavy quark fields transform
trivially under translations by one lattice unit.  Staggered bilinears
with non-singlet taste, 
on the other hand, get multiplied by phase factors under single-site translations. Therefore
mixed four quark operators cannot be translation singlets unless the light bilinear is
a singlet under taste.

The mixed four-quark operators are thus 
irrelevant from the point of view of our heavy-light chiral theory. Such operators are
invariant under the light quark chiral symmetries since they do not break taste (or, trivially,
flavor) symmetries and are independent of the light quark masses (otherwise, they would have
dimension greater than 6). We therefore can classify the discretization
errors caused by such operators as ``heavy-quark errors.''  These errors can
be estimated by the methods of Refs.~\cite{Kronfeld:2000ck,KRONFELD-REVIEW,OKAMOTO-05}.
(See especially the Appendix of Ref.\cite{OKAMOTO-05}
for estimates of heavy quark discretization errors
in a practical case.) In fact, one expects that errors from mixed four-quark operators
will be rather smaller than the more familiar ones from heavy-quark bilinear operators,
since the former are $\cO(\alpha_S^2a^2)$ for improved staggered (``Asqtad'') light quarks \cite{Asqtad}; 
while the latter are $\cO(\alpha_Sa)$ or $\cO(a^2)$.
It is important to keep in mind, however, that such error estimates are based on power
counting and dimensional analysis and must be considered as rough guides only.
Precise quantification of the total discretization error will always require simulation at several lattice spacings.

Four-quark operators built entirely out of heavy quarks are  
also invariant (trivially) under the light quark symmetries.  Therefore, the arguments of
the previous paragraph apply, with the exception that the discretization errors are
now  $\cO(\alpha_Sa^2)$: There is no suppression of the exchange of a single gluon with momentum
$\pi/a$ between the two quarks, unlike the cases where one or both of the quarks has the Asqtad action.

The relevant part of the Symanzik action may now be written 
as $S^{\rm Sk} = \int d^4x \cL^{\rm Sk}$, with
\begin{equation}\label{eq:LSymanzik}
  \cL^{\rm Sk}  =  \cL^{\rm Sk}_4 + a^2 \cL^{\rm Sk}_6 + \cO(a^4) \ ,
\end{equation}
where $\cL^{\rm Sk}_6 $ comes from light four-quark operators only.
$\cL^{\rm Sk}_4$ is the continuum limit of the lattice theory with
$n$ staggered light quarks (each with 4 tastes), and a single heavy quark. 
It has the form
\begin{equation}\label{eq:QCD}
   \cL^{\rm Sk}_4  =  \sum_{j=1}^n \bar{q}_j (i\Dslash - m_j) q_j
  + \overline{Q}(iv\cdot D)Q + \cL_{\rm gluons}\ .
\end{equation}
Here $v$ is the heavy quark velocity, and $j$ runs over
the  light quark flavors.
$\cL_{{\rm Sk},4} $ has a heavy quark $SU(2)$ spin symmetry and, 
in the limit that $m_j\to 0$, a 
chiral $SU(4n)_L\times SU(4n)_R$ symmetry.

The  $\cO(a^2)$ term $\cL^{\rm Sk}_6$ in \eq{LSymanzik} is
a sum over all 4-light-quark operators in the two classes $S_6^{FF(A)}$ and $S_6^{FF(B)}$
identified in Ref.~\cite{LEE_SHARPE}: 
\begin{equation}\label{eq:L6Symanzik}
\cL^{\rm Sk}_6 =  \cL^{\rm Sk}_{A} + \cL^{\rm Sk}_{B}\ . 
\end{equation}
$S_6^{FF(A)}$ consists of ten operators that do not violate
continuum rotation symmetry, while the four operators in
$S_6^{FF(B)}$ violate this symmetry.  For $n$ staggered fields, the operators in each of 
these classes are written down in Ref.~\cite{Sharpe:2004is}.
Every operator is
a product of two bilinears, each of which has the same spin and taste.  Denoting
the five possible spins structures by $S$, $V$, $T$, $A$, $P$, and similarly for tastes, the 
operators
are named by the spin $\otimes$ taste of their bilinears:
\begin{eqnarray}
\cL^{\rm Sk}_{A}: && [S\times A],\  [S\times V],\  [A\times S],\  [V\times S],\  [P\times A],\  [P\times V],\ 
 \nonumber \\
             && [A\times P],\  [V\times P],\  [T\times V],\  [T\times A],\  [V\times T],\  [A\times T]\ 
\label{eq:SFFA}
\\
\cL^{\rm Sk}_{B}:  && [T_\mu\times V_\mu],\  [T_\mu\times A_\mu],\  [V_\mu\times T_\mu],\  [A_\mu\times T_\mu]
\label{eq:SFFB}
\end{eqnarray}
In $\cL^{\rm Sk}_{A}$ the spin indices and taste indices are contracted separately, for example,
\begin{equation}
	[T\times A] \equiv \sum_{\mu<\nu} \sum_\lambda \bar q_i(\gamma_{\mu\nu} \otimes \xi_{\lambda5})q_i\;
\bar q_j(\gamma_{\nu\mu} \otimes \xi_{5\lambda})q_j
\end{equation}
where  $\gamma_{\mu\nu}\equiv (1/2) [\gamma_\mu,\gamma_\nu]$, $\gamma_{\mu5}\equiv \gamma_\mu\gamma_5$
(similarly for tastes, with $\gamma_\mu\to\xi_\mu$), and there are implicit sums over the flavor
indices $i,j$.
The operators in $\cL^{\rm Sk}_{B} $ are more complicated:
spin and taste indices are summed together, and two terms, with different treatment
of the tensor indices, are subtracted to cancel taste- or spin-singlet contributions. For example,
\begin{equation}
	[T_\mu\times A_\mu] \equiv\sum_\mu \sum_{\mu\not=\nu} \Big\{
\bar q_i\left(\gamma_{\mu\nu} \otimes \xi_{\mu5}\right)q_i\;
\bar q_j\left(\gamma_{\nu\mu} \otimes \xi_{5\mu}\right)q_j\;
-\bar q_i\left(\gamma_{\mu\nu5} \otimes \xi_{\mu5}\right)q_i
\bar q_j\left(\gamma_{5\nu\mu} \otimes \xi_{5\mu}\right)q_j\Big\}
\end{equation}
Because the same index ($\mu$) appears four times in each term in  $S_6^{FF(B)}$, these operators
are not invariant under continuum Euclidean rotations or taste transformations, 
but only under the lattice symmetry
where taste and Euclidean rotations by $90^\circ$ are performed simultaneously.

In Refs.~\cite{LEE_SHARPE} and \cite{SCHPT}, the leading order (LO) chiral Lagrangian for
light-light pseudoscalars is found for the one-flavor and many-flavor cases, respectively.
There the only available Lorentz 4-vector in the chiral theory is
the partial derivative operator $\partial_\mu$.  To construct a chiral operator with
the symmetries of $\cL^{\rm Sk}_{B}$ operators requires at least two derivatives and two explicit taste indices.
Such operators are $\cO(p^2a^2)$, where the factors of lattice spacing $a$ come from
the explicit violation of taste symmetry, and the factors of $p$ come from the derivatives. These
operators then only occur at next-to-leading-order (NLO) in the chiral Lagrangian, unlike
the representatives of $\cL^{\rm Sk}_{A}$ which are LO.  The $\cL^{\rm Sk}_{B}$ representatives are therefore
omitted in  Refs.~\cite{LEE_SHARPE,SCHPT}, although they are 
included in Ref~\cite{Sharpe:2004is}, which works
to NLO.

In the heavy-light case, however, there are two 4-vectors available in
the chiral theory whose presence does not raise the order of the operator: 
 the heavy quark 4-velocity $v_\mu$ and the light quark gamma matrix $\gamma_\mu$.
Therefore the chiral representatives of the $\cL^{\rm Sk}_{A}$  operators contribute at the
same order as the representatives of the $\cL^{\rm Sk}_{B}$.
For heavy-light decay constants this is NLO --- such
operators enter into analytic terms in the calculation presented here, but not in one-loop
diagrams.

\section{The \schpt\ Lagrangian with Heavy-Light Mesons}
\label{sec:ls-lag}

\subsection{Leading order continuum theory}
\label{sec:LOcont}

We first discuss standard \chpt\ at LO for
heavy-light mesons in 
the continuum \cite{Burdman:1992gh,Grinstein-et,Goity:1992tp,MAN_WISE}. 
More precisely in the current development, 
this is the LO chiral representative of the dimension
4 part of the Symanzik Lagrangian, $\cL^{\rm Sk}_4$, \eq{QCD}.
Due to
the heavy quark spin symmetry in the static limit, the heavy vector
and pseudoscalar mesons are incorporated into the following field,
which destroys a heavy-light meson
\begin{equation}
  H_a = \frac{1 + \vslash}{2}\left[ \gamma^\mu B^{*}_{\mu a}
    + i \gamma_5 B_{a}\right]\ ,
\end{equation}
where $v$ is the meson's velocity, and $a$ is the combined light quark flavor-taste index. 
The conjugate field creates a heavy-light meson
\begin{equation}
  \overline{H}_a \equiv \gamma_0 H^{\dagger}_a\gamma_0 =
  \left[ \gamma^\mu B^{\dagger *}_{\mu a}
    + i \gamma_5 B^{\dagger}_{a}\right]\frac{1 + \vslash}{2}\ ,
\end{equation}
We use $B$ to
denote a generic pseudoscalar heavy meson and $B^*$ to denote the corresponding
vector meson, but note that the current formalism will also apply to $D$ and $D^*$ mesons,
although at a decreased level of accuracy in the $1/m_Q$ expansion.

Under the $SU(2)$ heavy quark spin symmetry, the heavy-light field
transforms as
\begin{eqnarray}
  H &\to & S H\ , \nonumber\\
  \overline{H} &\to & \overline{H}S^{\dagger}\ ,\label{eq:Hspin}
\end{eqnarray}
with $S\in SU(2)$, while under
the $SU(4n)_L\times SU(4n)_R$ chiral symmetry,
\begin{eqnarray}
  H &\to & H \mathbb{U}^{\dagger}\ ,\nonumber\\
  \overline{H} &\to & \mathbb{U}\overline{H}\ ,
\end{eqnarray}
with $\mathbb{U}\in SU(4n)$, and we are keeping the flavor-taste
 index implicit. 

We use the term ``pion'' 
generically to denote any of the light pseudoscalar mesons. 
The field describing the pions is 
$\Sigma = \exp [i\Phi/f]$, where
$f$ is the tree-level pion decay constant ($f \sim f_\pi \cong 131\ \MeV$). 
In terms involving the heavy-lights, we also need $\sigma \equiv
\sqrt{\Sigma} = \exp[ i\Phi / 2f ]$. These fields are singlets under
the heavy-quark spin symmetry, while under $SU(4n)_L\times SU(4n)_R$ we have
\begin{eqnarray}
  \Sigma \to  L\Sigma R^{\dagger}\,,\qquad&&\qquad
  \Sigma^\dagger \to  R\Sigma^\dagger L^{\dagger}\,,\\*
  \sigma \to  L\sigma \mathbb{U}^{\dagger} = \mathbb{U} \sigma R^{\dagger}\,, \qquad&&\qquad
  \sigma^\dagger \to R \sigma^\dagger \mathbb{U}^{\dagger} = \mathbb{U} \sigma^\dagger L^{\dagger}\,, 
  \label{eq:Udef}
\end{eqnarray}
with $L\in SU(4n)_L$ and $R\in SU(4n)_R$. \Eq{Udef} defines $\mathbb{U}$, which is a function of the pion fields and the coordinates, 
as well as of the global transformations $L$ and $R$.
Chiral operators are formed from combinations of $H$, $\overline{H}$, $\sigma$, $\sigma^\dagger$
and derivatives, as well other matrices such as the light quark mass matrix or taste matrices,
so that they transform in the same manner under the
combined symmetry group as the underlying operators in the Symanzik
action.  As usual, this is accomplished by promoting any symmetry-violating factors
in the Symanzik
Lagrangian to spurions and choosing the transformation properties of the spurions
to make the Lagrangian invariant.

For $n$ KS flavors, $\Sigma=\exp(i\Phi / f)$ is a $4n \times 4n$
matrix, and $\Phi$ is given by:
\begin{eqnarray}\label{eq:Phi}
  \Phi = \left( \begin{array}{cccc}
      U  & \pi^+ & K^+ & \cdots \\*
      \pi^- & D & K^0  & \cdots \\*
      K^-  & \bar{K^0}  & S  & \cdots \\*
      \vdots & \vdots & \vdots & \ddots \end{array} \right),
\end{eqnarray}
where $U = \sum_{\Xi=1}^{16} U_\Xi T_\Xi$, \etc, with the Hermitian taste
generators $T_\Xi$ given by
\begin{equation}\label{eq:T_Xi}
  T_\Xi = \{ \xi_5, i\xi_{\mu 5},
  i\xi_{\mu\nu} , \xi_{\mu},
  \xi_I\}\ .
\end{equation}
As in Ref.~\cite{SCHPT} the
taste matrices are $\xi_{\mu}$, with 
$\xi_{\mu5}\equiv \xi_{\mu}\xi_5$. 
The $4\times 4$ identity matrix is $\xi_I \equiv I$.
In this paper, we define
$\xi_{\mu\nu}\equiv (1/2)[\xi_{\mu},\xi_{\nu}]$ (rather than simply
$\xi_{\mu}\xi_{\nu}$) to emphasize that terms with $\mu=\nu$ are never 
included; in 
\eq{T_Xi} we take 
$\mu <\nu$ only.

The component fields of the diagonal (flavor-neutral)
elements ($U_\Xi$, $D_\Xi$, \etc) are real; the other (charged)
fields are complex ($\pi^+_\Xi$, $K^0_\Xi$, \etc), such that $\Phi$ is
Hermitian. Here the $n=3$ portion of $\Phi$ is shown explicitly.  The
mass matrix is the $4n\times 4n$ matrix
\begin{eqnarray}
  \cM = \left( \begin{array}{cccc}
      m_u I  & 0 &0  & \cdots \\*
      0  & m_d I & 0  & \cdots \\*
      0  & 0  & m_s I  & \cdots\\*
      \vdots & \vdots & \vdots & \ddots \end{array} \right),
\end{eqnarray}
where again, the portion shown is for the $n=3$ case.

It is convenient to define objects involving the $\sigma$ field that transform 
only with $\mathbb{U}$ and $\mathbb{U}^\dagger$. The two 
possibilities with a single derivative are
\begin{eqnarray}
  \mathbb{V}_{\mu} & = & \frac{i}{2} \left[ \sigma^{\dagger} \partial_\mu
   \sigma + \sigma \partial_\mu \sigma^{\dagger}   \right] \ ,\\
  \mathbb{A}_{\mu} & = & \frac{i}{2} \left[ \sigma^{\dagger} \partial_\mu
   \sigma - \sigma \partial_\mu \sigma^{\dagger}   \right] \ ,
\end{eqnarray}
$\mathbb{V}_{\mu}$ transforms like a vector field under the $SU(4n)_L\times SU(4n)_R$ 
chiral symmetry and, when combined with the derivative, can form a 
covariant derivative acting on the heavy-light field or its conjugate: 
\begin{eqnarray}\label{eq:Ddef}
(H \leftvec D_\mu)_a  = H_b \leftvec D^{ba}_\mu  &\equiv& \partial_\mu H_a + i H_b\mathbb{V}_{\mu}^{ba}\ , \nonumber \\
(\rightvec D_\mu \overline{H})_a  = \rightvec D^{ab}_\mu \overline{H}_b &\equiv& \partial_\mu \overline{H}_a - 
i \mathbb{V}_{\mu}^{ab} \overline {H}_b\ ,
\end{eqnarray}
with implicit sums over repeated indices.
The covariant derivatives and $\mathbb{A}_\mu$
transform under the chiral symmetry as
\begin{eqnarray}\label{eq:Dtransf}
	H \leftvec D_\mu &\to&  (H \leftvec D_\mu )\mathbb{U}^\dagger\ , \nonumber \\
	\rightvec D_\mu \overline{H} &\to&  \mathbb{U} (\rightvec D_\mu \overline{H})\ ,\nonumber \\
	 \mathbb{A}_\mu &\to&  \mathbb{U} \mathbb{A}_\mu \mathbb{U}^\dagger\ .
\end{eqnarray}

We then write the (Minkowski space) continuum LO chiral Lagrangian
as 
\begin{equation}\label{eq:Lcont}
  \cL_{\rm LO,cont} = \cL_{\rm LO,cont}^{\rm pion}+ \cL_1 \ ,
\end{equation}
where $\cL_{\rm LO,cont}^{\rm pion}$ is the standard 
LO light meson Lagrangian in the continuum, and $ \cL_1$
is the leading term involving the heavy-lights. As explained in \secref{power},
lattice corrections affect the heavy-light terms only at NLO, so there is no
need to specify ``cont'' on $ \cL_1$.
We have
\begin{equation}\label{eq:Lcontpion}
	\cL_{\rm LO,cont}^{\rm pion} = \frac{f^2}{8} \Tr(\partial_{\mu}\Sigma 
  \partial^{\mu}\Sigma^{\dagger}) + 
  \frac{1}{4}\mu f^2 \Tr(\cM\Sigma+\cM\Sigma^{\dagger})
  - \frac{2m_0^2}{3}(U_I + D_I + S_I+\ldots)^2 
\end{equation}
and \cite{MAN_WISE}
\begin{equation}\label{eq:L1}
  \cL_1  =  -i \Tr(\overline{H} H v\negcdot \leftvec D )
  + g_\pi \Tr(\overline{H}H\gamma^{\mu}\gamma_5 
  \mathbb{A}_{\mu}) \ .
\end{equation}
$\Tr$ means the complete trace over flavor-taste indices and, where relevant,
Dirac indices. Since $\overline H$ and $H$ always appear together  in the Lagrangian,
we treat $\overline{H}H$ as a matrix in flavor-taste space: 
$(\overline{H}H)_{ab} \equiv \overline{H}_aH_b$.
Here and below, the covariant 
derivative $\leftvec D$ acts only on the field (in this
case $H$) immediately 
preceding it; similarly $\rightvec D$ acts only on the field immediately following it.
As in Ref.~\cite{SCHPT}, it is helpful for analyzing the quark
flow to leave the anomaly ($m_0^2$) term explicit
in \eq{Lcontpion} and work with the diagonal fields $U$, $D$, $S$, \dots, rather
than the physical ones $\pi_0$, $\eta$, \dots.  At the end of the calculation,
we can take $m^2_0\to\infty$ and return to the physical  basis \cite{SHARPE_SHORESH}.

Although fundamentally we are interested in  the chiral effective theory
for a {\it Euclidean} lattice theory, we find it more convenient to write 
the Lagrangian in Minkowski space to make direct contact with the extensive
continuum literature.  A Wick rotation to define the Euclidean
Green's functions is implicit everywhere.

For the heavy-light decay constants, we also need the chiral representative 
of the axial heavy-light current.  Alternatively, one can work with the
left-handed current, which has the advantage that it transforms more simply under
chiral transformations and in addition  gives corresponding information about the
vector current, useful for semileptonic form factors.
The left-handed current that destroys a heavy-light meson
of flavor-taste $b$ is $j^{\mu,b}$, which at LO takes the form 
\cite{MAN_WISE} 
\begin{equation}\label{eq:LOcurrent}
  j^{\mu,b}_{\rm LO} = \frac{\kappa}{2}\; 
  \trD\bigl(\gamma^\mu \left(1-\gamma_5\right) H\bigr) \sigma^\dagger \lambda^{(b)}
\end{equation}
where $\lambda^{(b)}$ is a constant vector that fixes the flavor-taste:
$(\lambda^{(b)})_a = \delta_{ab}$, and $\trD$ is a trace on Dirac indices only.  
In QCD, the decay constant $f_{B_a}$ is defined by the matrix
element
\begin{equation}\label{eq:matrix_element}
  \left\langle 0 \left| j^{\mu,b}
  \right| B_{a}(v) \right\rangle =i
  f_{B_{a}}m_{B_{a}} v^{\mu} \delta_{ab}\ ,
\end{equation}
where relativistic normalization of the state $|B_{a}(v) \rangle $ is assumed.
At LO in the heavy-light chiral theory,
$j^{\mu,b}_{\rm LO} = i\kappa v^\mu {B_b}$, which gives 
$f_{B_{a}}^{\rm LO} = \kappa/\sqrt{m_{B_{a}}}$. Recall that the
factor $\sqrt{m_{B_{a}}}$ arises from the differences in
normalizations between relativistic and non-relativistic states.

\subsection{Power counting}
\label{sec:power}

Before considering discretization errors and higher order 
corrections to \eqs{Lcont}{LOcurrent},
we discuss the power counting assumed in this paper.
The staggered chiral Lagrangian with only light mesons is a joint expansion in the
light quark mass $m_q $ and the lattice spacing $a^2$. 
In that case, $m_q$ and $a^2$ are taken to be of the same order in the 
expansion \cite{LEE_SHARPE,SCHPT}. 
Since $m_\pi^2\propto m_q$ and
the momentum of external light mesons, $p_\pi$, is assumed to be of order $m_\pi$,
we have $p_\pi^2\sim m^2_\pi\sim m_q\sim a^2$ in the power counting.

Upon including $B$ mesons, we have an additional
expansions in the inverse of the heavy quark mass $m_Q$, and in
the $B$ meson's residual momentum, $k$,
which we take to be of the same order as $p_\pi$.  
We work to leading order in $1/m_Q$ only.

The continuum LO chiral Lagrangian, \eq{Lcont}, 
is therefore $\cO(k\sim \sqrt{m_q})$ in the heavy-meson fields
and $\cO(m_q,a^2)$ in the light meson fields. In each case these are the lowest order
terms allowed by the continuum symmetries.\footnote{The $\cO(k^0)$ heavy quark mass term is absorbed
by measuring heavy-light energies relative to the mass of the heavy meson.} 
One-loop diagrams with vertices from \eqs{Lcont}{LOcurrent} and propagators
from \eq{Lcont} will give $\cO(m_\pi^2)$ corrections to physical
quantities such as $f_{B_a}$ or semileptonic form factors. 
Since taste-violating terms of $\cO(a^2)$  in the pion
Lagrangian of \schpt\ are the same order as those already included in \eq{Lcontpion},
we will need to include such terms in order to calculate the one-loop corrections.

Terms that are higher order than those in
\eqs{Lcont}{LOcurrent} are irrelevant in the one-loop diagrams.  They can, however,
make analytic (tree-level) contributions at this (or, in principle, lower) order.  
In particular, for $f_{B_a}$ we will need
to consider $\cO(\sqrt{m_q})$ and $\cO(m_q)$ corrections to the existing
heavy-light terms in \eqs{Lcont}{LOcurrent}.
Such terms in the heavy-light Lagrangian, \eq{L1} are of
order $m_q\sim a^2$ and $km_q\sim ka^2$; those in the current, \eq{LOcurrent},
are of order $m_q\sim a^2$.  These corrections are generated by inserting, into heavy-light chiral
operators, spurions representing the light-quark mass term in $\cL^{\rm Sk}_4$ [\eq{QCD}] and
the $\cO(a^2)$ taste-violating terms, $\cL^{\rm Sk}_6$ [\eq{L6Symanzik}].
There are also invariant higher order operators that  are present
even in the continuum and chiral limits.  These are
generated by inserting additional derivative
operators, \ie additional powers of $k$.
As explained in \secref{HigherDerivs}, such terms do not affect the decay constants to the order
we are working, and we do not attempt to catalog them completely.  
We also do not need to consider here higher order
[$\cO(m_q^2\sim m_qa^2\sim a^4)$] corrections to the pion Lagrangian. Such corrections
cannot give analytic contributions to 
heavy-light decay constants.  They are, however, relevant to purely light-light physics,
as well as to semileptonic form factors of heavy-light mesons, which involve a light-light
particle.  All such terms have been found in Ref.~\cite{Sharpe:2004is}.

Summarizing the discussion of this subsection, we write our complete chiral Lagrangian as
\begin{eqnarray}
\cL & = & \cL_{\rm LO} + \cL_{\rm NLO}\ ,\label{eq:Lcomplete}\\
\cL_{\rm LO} & = & \cL_{\rm LO}^{\rm pions} +\cL_1=\cL_{\rm LO,cont}^{\rm pions} 
+ \cL_{{\rm LO,}a^2}^{\rm pions} +\cL_1 \ ,\label{eq:LLO}\\
\cL_{\rm NLO} & = & \cL_2 + \cL_3 \, \label{eq:LNLO}\\
\cL_2 & = & \cL_{2,k} + \cL_{2,m} + \cL_{2,a^2}^{A} +  \cL_{2,a^2}^{B} \ , \label{eq:L2}\\
\cL_3 & = & \cL_{3,k} + \cL_{3,m} + \cL_{3,a^2}^{A} +  \cL_{3,a^2}^{B} \ . \label{eq:L3}
\end{eqnarray}
Here, $\cL_1$, 
$\cL_2$, and $\cL_3$ involve the heavy-light fields and are of order $k$, $k^2\!\sim\! m_q\!\sim\! a^2$, and
$k^3\!\sim\! km_q \!\sim\! ka^2$, respectively.  
The subscripts $k$, $m$, or $a^2$ indicate terms that involve 
derivatives only, mass spurions, or taste-violating spurions. The superscripts
$A$ and $B$ stand for the type of taste-violating terms, \eqs{SFFA}{SFFB}.

Similarly we have for the left-handed current:
\begin{eqnarray}
j^{\mu,b} & = & j^{\mu,b}_{\rm LO} + j^{\mu,b}_{\rm NLO}\ ,\label{eq:completecurrent}\\
j^{\mu,b}_{\rm NLO} & = & j^{\mu,b}_1 + j^{\mu,b}_2 \, \label{eq:NLOcurrent}\\
j^{\mu,b}_1 & = & j^{\mu,b}_{1,k} \label{eq:current1} \\
j^{\mu,b}_2 & = & j^{\mu,b}_{2,k} + j^{\mu,b}_{2,m} + j^{\mu,b}_{2,a^2,A} +  j^{\mu,b}_{2,a^2,B} \ . \label{eq:current2} 
\end{eqnarray}
Note that insertions of the mass and $a^2$ spurions are always at least two orders higher in
$k\!\sim\!\sqrt{m_q}$, so do not appear in $j^{\mu,b}_1 $.

The continuum LO  terms $\cL_1$, $\cL_{\rm LO, cont}^{\rm pions} $, and $ j^{\mu,b}_{\rm LO} $ are given 
in \eqsthree{Lcontpion}{L1}{LOcurrent}.
The remaining contributions, Lagrangian terms $\cL_{{\rm LO,}a^2}^{\rm pions} $, $\cL_2$, and $\cL_3$,
and current terms $j^{\mu,b}_1 $ and $j^{\mu,b}_2 $, are discussed in the following subsections.

\subsection{LO discretization effects}
\label{sec:LOa2}

The LO finite lattice spacing correction, $\cL_{{\rm LO,}a^2}^{\rm pion}$, to the
pion Lagrangian is known from previous work \cite{LEE_SHARPE,SCHPT}.
We have 
\begin{eqnarray}
	\cL_{{\rm LO,}a^2}^{\rm pion} &=& - a^2\cV_\Sigma  \nonumber\\
   \cV_\Sigma & = & C_1
  \Tr(\xi^{(n)}_5\Sigma\xi^{(n)}_5\Sigma^{\dagger})
  +C_3\frac{1}{2} \sum_{\nu}[ \Tr(\xi^{(n)}_{\nu}\Sigma
    \xi^{(n)}_{\nu}\Sigma) + h.c.] \nonumber \\*&&
  {}+C_4\frac{1}{2} \sum_{\nu}[ \Tr(\xi^{(n)}_{\nu 5}\Sigma
    \xi^{(n)}_{5\nu}\Sigma) + h.c.]
   +C_6\ \sum_{\mu<\nu} \Tr(\xi^{(n)}_{\mu\nu}\Sigma
  \xi^{(n)}_{\nu\mu}\Sigma^{\dagger})  \nonumber \\*&&
   {}+C_{2V}\frac{1}{4} \sum_{\nu}[ \Tr(\xi^{(n)}_{\nu}\Sigma)
   \Tr(\xi^{(n)}_{\nu}\Sigma)
    + h.c.]
   +C_{2A}\frac{1}{4} \sum_{\nu}[ \Tr(\xi^{(n)}_{\nu5}\Sigma)
   \Tr(\xi^{(n)}_{5\nu}\Sigma)
    + h.c.] \nonumber \\*
  && {}+C_{5V}\frac{1}{2} \sum_{\nu} \Tr(\xi^{(n)}_{\nu}\Sigma)
  \Tr(\xi^{(n)}_{\nu}\Sigma^{\dagger})
  + C_{5A}\frac{1}{2}\sum_{\nu}\Tr(\xi^{(n)}_{\nu 5}\Sigma)
  \Tr(\xi^{(n)}_{5\nu}\Sigma^{\dagger}) \label{eq:VSigma} \ .
\end{eqnarray}
The explicit $4n\times 4n$ matrices
$\xi^{(n)}_{\mu}$ in \eq{VSigma} are defined by
\begin{equation}\label{eq:xi-n}
  \left(\xi^{(n)}_{\nu}\right)_{ij} = \xi_{\nu}\delta_{ij}\ ,
\end{equation}
with $i$ and $j$ the $SU(n)$ (light quark) flavor indices,
and $\xi_\nu$ the $4\times4$ taste matrix, as in \eq{T_Xi}.
   The matrices $\xi^{(n)}_{\mu\nu}$ and 
   $\xi^{(n)}_{\nu5}$ are defined similarly. 

There are no $a^2$ corrections at LO
to $\cL_1$, \eq{L1},  or to the left-handed current, \eq{LOcurrent}.

\subsection{Invariant NLO corrections; $\cT$ symmetry}
\label{sec:HigherDerivs}

Even in the chiral and continuum limit, there are invariant terms that
can be constructed by adding more derivatives to the lowest-order heavy-light
Lagrangian $\cL_1$, or current $j^{\mu,b}_{\rm LO}$.  To our knowledge, 
a complete catalog of such terms
through $\cO(k^3)$ does not appear in the continuum literature. 
These higher derivative terms are in fact irrelevant for the heavy-light
decay constant to the order we are working:  Derivatives acting on
the light-light fields $\Sigma$ or $\sigma$  do not contribute to
tree diagrams
because there are no external pions in the matrix element; while terms
with a derivative acting on the heavy-light field in the current,
or more than one derivative acting on a heavy-light field in the Lagrangian, give a
vanishing contribution on shell ($k=0$).\footnote{Terms with a single derivative in
the Lagrangian do contribute on shell through wave-function renormalization.}

For semileptonic form factors, \eg those for $B\to\pi$,
higher derivative terms do produce analytic corrections at this
order.
However, the functional form that such contributions can take
is rather limited:  In the chiral limit, one 
can only get corrections to form
factors proportional to $v\cdot p$ or $(v\cdot p)^2$, where 
$p$ is the pion 4-momentum.
Away from the chiral limit, there will be additional corrections
proportional to $p^2$, but these merely duplicate, on-shell, the effects
of mass or $a^2$ spurions.

We thus simply list here a few representative operators that appear in
$\cL_{2,k}$, $\cL_{3,k}$, $j^{\mu,b}_{1,k}$, and $j^{\mu,b}_{2,k}$:
\begin{eqnarray}
\cL_{2,k} &=& \frac{i\epsilon_1}{\Lambda_\chi}\Tr\left((v\cdot \rightvec D\, \overline{H}H 
- \overline{H}H v\cdot\leftvec D)\,\gamma_\mu \gamma_5 \mathbb{A}^\mu\right)
+ \frac{\epsilon_2}{\Lambda_\chi}\Tr\left(\overline{H} (v\cdot \rightvec D\,)^2 H\right)   + \dots 
\label{eq:L2k} \\
\cL_{3,k} &=& 
\frac{\epsilon_3}{\Lambda_\chi^2}\Tr\left(\overline{H}H \gamma_\mu \gamma_5 
(v\cdot \rightvec D\,)^2 \mathbb{A}^\mu\right) 
+ \frac{\epsilon_4}{\Lambda_\chi^2}\Tr\left(\overline{H} H \rightvec \Dslash\,\gamma_5    \, v\cdot \rightvec D \, 
 v\cdot \mathbb{A}\right) + \dots 
\label{eq:L3k} \\
  j^{\mu,b}_{1,k} &=& \frac{i\kappa_1}{\Lambda_\chi}\; 
  \trD\bigl(\gamma^\mu \left(1\!-\!\gamma_5\right) v\cdot \rightvec D\, H\bigr) 
\sigma^\dagger \lambda^{(b)} 
   + \frac{\kappa_2}{\Lambda_\chi}\; 
  \trD\bigl(\gamma^\mu \left(1\!-\!\gamma_5\right) H\,\bigr) 
v\cdot  \mathbb{A}\, \sigma^\dagger \lambda^{(b)} +\dots 
\label{eq:j1k} \\
  j^{\mu,b}_{2,k} &=& \frac{\kappa_3}{\Lambda^2_\chi}\; 
  \trD\bigl(\gamma^\mu \left(1\!-\!\gamma_5\right) (v\cdot \rightvec D\,)^2 H\bigr) 
\sigma^\dagger \lambda^{(b)} \nonumber 
\label{eq:j2k} \\
   &&\quad + \frac{i\kappa_4}{\Lambda^2_\chi}\; 
  \trD\bigl(\gamma^\mu \left(1\!-\!\gamma_5\right) H\,\bigr) 
v\cdot \rightvec D\, v\cdot  \mathbb{A}\, \sigma^\dagger \lambda^{(b)} +\dots
\end{eqnarray}
where the constants $\epsilon_i, \kappa_j$ are taken to be real and dimensionless, 
$\Lambda_\chi$ is the chiral scale, and
\begin{equation}
\rightvec D_\nu \mathbb{A}_\mu \equiv \partial_\nu \mathbb{A}_\mu   -i [\mathbb{V}_\nu, \mathbb{A}_\mu] \ .
\end{equation}

The use of time reversal ($\cT$) symmetry and the requirement of Hermiticity
for the Lagrangian can be used to eliminate several other candidate operators 
from \eqsthru{L2k}{j2k}.  Because we will also need to use the consequences of $\cT$
invariance extensively in \secthru{NLOm}{NLOa2B}, we briefly review how it
acts on relevant objects, following Ref.~\cite{BOYD}.  Let $\tilde p$ be the
parity reflection of a 4-vector $p$; in other words $\tilde p^\mu=p_\mu$.
Then time reversal acts as follows
\begin{eqnarray}
x &\to& -\tilde x \nonumber \\
v &\to& \tilde v \nonumber \\
\Phi(x) &\to& -\Phi(-\tilde x) \nonumber \\
\Sigma(x) &\to& \Sigma(-\tilde x) \nonumber \\
\sigma(x) &\to& \sigma(-\tilde x) \nonumber \\
\mathbb{A}^\mu(x) &\to& \mathbb{A}_\mu(-\tilde x) \nonumber \\
\mathbb{V}^\mu(x) &\to& \mathbb{V}_\mu(-\tilde x) \nonumber \\
\rightvec D^\mu  &\to& -\rightvec D_\mu \nonumber \\
H^v(x) &\to& T\, H^{\tilde v}(-\tilde x) \,T^{-1} \nonumber \\
\overline H^{\,v}(x) &\to& T\, \overline H^{\,\tilde v}(-\tilde x) \,T^{-1} \label{eq:Tinv}
\end{eqnarray}
where the Dirac matrix $T$ has the property $T\gamma^\mu T^{-1}=\gamma_\mu^*$,
we show the $v$ dependence of $H$ explicitly, 
and we have used the anti-unitary nature of $\cT$.  

Requiring $\cT$ invariance, we can for example eliminate contributions to the Lagrangian such as
\begin{equation}
\Tr\left(\overline{H}H \gamma_\mu \gamma_5 
v\cdot \rightvec D\, \mathbb{A}^\mu\right) \ .
\end{equation}
Adding a factor of $i$ to make this $\cT$-even, as was done 
in Ref.~\cite{BOYD}, does not in fact save this term because it is then
anti-Hermitian.

\subsection{Quark mass corrections at NLO}
\label{sec:NLOm}

Here we discuss operators induced by single insertions of the light quark mass
spurions in $\cL^{Sk}_4$. These produce corrections to the Lagrangian, namely
$\cL_{2,m}$ and $\cL_{3,m}$, and corrections to the current, $j^{\mu,b}_{2,m}$.
Of course, such terms are also present at NLO in the continuum, and the result
is more or less standard \cite{BOYD,Stewart:1998ke}.  However, since many of the
contributions to the discretization corrections 
($\cL_{2,a^2}^A$,$\cL_{2,a^2}^B$, $\cL_{3,a^2}^A$, $\cL_{3,a^2}^B$, $j^{\mu,b}_{2,a^2,A}$, and 
$j^{\mu,b}_{2,a^2,B}$)
follow the same pattern (see \secrefs{NLOa2A}{NLOa2B}), it
is worthwhile to examine the mass corrections carefully first. In addition, 
we find some modifications to the terms written down in Refs.~\cite{BOYD,Stewart:1998ke}.

We start by writing down combinations of the light quark mass spurions
that transform  only with $\mathbb{U}$ or $\mathbb{U}^\dagger$. These are
\begin{equation}\label{eq:Mpm}
  \cM^\pm  =  \frac{1}{2}\left(\sigma \cM\sigma
  \pm \sigma^{\dagger} \cM\sigma^{\dagger}\right) \ ,
\end{equation}
which transform as $\cM^{\pm}\to \mathbb{U} \cM^{\pm} \mathbb{U}^\dagger$ under the
chiral symmetry.  
Under parity, $\cM^\pm \to \pm \cM^\pm$, since $\Phi\to-\Phi$ and
$\sigma\to\sigma^\dagger$; while both  $\cM^{+}$ and  $\cM^{-}$ are even
under time reversal.   Because $\trD (\overline{H}H\gamma_5)=0$, the only
parity and chiral invariants we can construct with no derivatives are
$\Tr(\overline{H} H\cM^+)$ and $\Tr(\overline{H}H)\Tr(\cM^+)$.  
We thus have
\begin{equation}\label{eq:L2m}
  \cL_{2,m}  =  2\lambda_1 \Tr\left(\overline{H} H\cM^+\right) 
  + 2\lambda'_1 \Tr\left(\overline{H} H\right)\Tr\left(\cM^+\right)   \ .
\end{equation}

For $ \cL_{3,m}$, we start with the lists of terms given 
in Eq.~(16) of Ref.~\cite{BOYD} and Eq.~(7) of Ref.~\cite{Stewart:1998ke}.
Note, however, that a term like
$\Tr\left( \overline{H} H  \gamma_\mu\gamma_5\,\mathbb{A}^\mu 
\cM^+\right)$ is not Hermitian, because the Hermitian conjugate will flip 
the order of $\cM^+$ and $\mathbb{A}^\mu$. To make it Hermitian we 
can write it in one of two ways
\begin{eqnarray}
	\Tr\left( \overline{H} H  \gamma_\mu\gamma_5\{\mathbb{A}^\mu, 
	\cM^+\} \right) \ ,\nonumber\\
	i\Tr\left( \overline{H} H  \gamma_\mu\gamma_5[\mathbb{A}^\mu, 
	\cM^+]\right)\ . \nonumber
\end{eqnarray}
The second term is not invariant under time-reversal and can be dropped.
For corresponding terms with $\cM^-$, it is the anticommutator that requires a
factor of $i$, since $\cM^-$ is anti-Hermitian. Then
time-reversal invariance requires that we build the traces using the commutator 
$[\mathbb{A}^\mu, \cM^-]$ only. 
Terms with separately traced 
operators, such as $\Tr\left( \overline{H} H  \gamma_\mu\gamma_5)\Tr(\mathbb{A}^\mu 
\cM^+\right)$, do not need to be rewritten with commutators or
anticommutators since they are Hermitian already. 

Terms involving a derivative acting on the heavy-light fields also require some
thought. Hermitian combinations are
\begin{eqnarray}
&i\left(\rightvec D_\mu \overline{H}H - \overline{H}H\leftvec D_\mu\right)\ , & \nonumber  \\
&\rightvec D_\mu \overline{H}H + \overline{H}H\leftvec D_\mu \ . & \nonumber  
\end{eqnarray}
However, we will not need to include
the second combination, which is a total derivative.  When combined  a factor
of $\cM^\pm$, integration by parts can be used to put the derivative on the
$\sigma$ or $\sigma^\dagger$ fields.  Such terms can then be rewritten in terms of
$\mathbb{A}$, using $i\rightvec D^{\,\mu}\sigma^\dagger=
-\mathbb{A}^\mu \sigma^\dagger$ or  $i\rightvec D^{\,\mu}\sigma=
\mathbb{A}^\mu \sigma$ \cite{BOYD}.

Putting together the above discussion, we have
\begin{eqnarray}\label{eq:L3m}
  \cL_{3,m} & = & 
i k_1 \Tr\left( \overline{H}H  v\negcdot \leftvec D\, \cM^+ - v\negcdot \rightvec D \,
\overline{H}H\, \cM^+
  \right) \nonumber \\
&&+ ik_2 \Tr\left( \overline{H} H  v\negcdot \leftvec D -v\negcdot \rightvec D \,
\overline{H}H\right)
  \Tr(\cM^+) \nonumber\\
  &&{}+ 
  k_3 \Tr\left( \overline{H} 
     H  \gamma_\mu\gamma_5\{\mathbb{A}^\mu ,
  \cM^+\}\right) +
  k_4 \Tr\left( \overline{H}
     H \gamma_\mu\gamma_5 \mathbb{A}^\mu \right)  
     \Tr(\cM^+)
   \nonumber\\
  &&{} + k_5 \Tr\left( \overline{H} 
     H  \gamma_\mu\gamma_5\right)\Tr\left(\mathbb{A}^\mu 
  \cM^+\right) +
  k_6 \Tr\left( \overline{H}
     H \gamma_\mu [\mathbb{A}^\mu ,
  \cM^-] \right) \ .
\end{eqnarray}
Note that $\overline{H}$ and $H$ must be next to each 
other in these traces by heavy quark spin symmetry.
The term that involves $\cM^-$ will not contribute to the decay 
constants or to the form factors for semileptonic decays, but we
include it here to complete the list of mass-dependent terms.

For the current, we have:
\begin{eqnarray}\label{eq:j2m}
  j^{\mu,b}_{2,m} & = & 
  \rho_1\,
  \trD\left(\gamma^\mu (1-\gamma_5) H  \right)
  \cM^+ \sigma^\dagger \lambda^{(b)}
  + \rho_2\,
  \trD\left(\gamma^\mu (1-\gamma_5) H\right) \sigma^\dagger
   \lambda^{(b)} 
  \Tr(\cM^+) \nonumber \\
&&+
\rho_3\,
\trD\left(\gamma^\mu (1-\gamma_5) H\right)  
\cM^- \sigma^\dagger \lambda^{(b)}
+ \rho_4\,
\trD\left(\gamma^\mu (1-\gamma_5) H\right) \sigma^\dagger
 \lambda^{(b)}
\Tr(\cM^-) \ .
\end{eqnarray}
It is not hard to argue that the heavy quark spin symmetry 
forces the factor $\gamma^\mu (1-\gamma_5) $
to appear next to the field $H$; it could not, for example, be replaced with 
$v^\mu (1-\gamma_5)$.  To show this, one can replace 
$\gamma^\mu (1-\gamma_5)$ in the QCD current by a spurion, and give it 
appropriate transformation properties under heavy quark spin symmetry 
and Lorentz transformations to make the current invariant.
A similar argument will apply even for the more complicated terms in the current
described  below.

\section{Taste breaking with Heavy-Light Mesons}\label{sec:4quark-ops}

The final needed contributions to the chiral Lagrangian and current are
$\cO(a^2)$ terms involving heavy-light fields: 
$\cL_{2,a^2}^{A}$, $\cL_{2,a^2}^{B}$, $\cL_{3,a^2}^{A}$, $\cL_{3,a^2}^{B}$ [\eqs{L2}{L3}] and
$j^{\mu,b}_{2,a^2,A}$, $j^{\mu,b}_{2,a^2,B}$ [\eq{current2}]. These are
constructed by starting with the dimension-6 terms in the Symanzik Lagrangian,
$\cL^{\rm Sk}_A$ and $\cL^{\rm Sk}_B$ [\eqs{SFFA}{SFFB}].
We promote the two explicit taste matrices in each of these operators
to spurion fields, which are assigned  transformation laws such that the operators
are invariant under the chiral $SU(4n)_L \times SU(4n)_R$ symmetry. 

The heavy quark spin symmetry requires that we only have the
combination $\overline{H} H$ in the Lagrangian, and $\gamma^\mu(1-\gamma_5)H$
in the current. Under the chiral symmetry, these transform as
\begin{eqnarray*}
  \overline{H} H &\to &\mathbb{U} \overline{H} H \mathbb{U}^{\dagger}\\
  \gamma^\mu(1-\gamma_5)H &\to & \gamma^\mu(1-\gamma_5) H \mathbb{U}^{\dagger}\ .
\end{eqnarray*}
When doing the spurion analysis,
we therefore need to combine the taste matrices (treated as spurion
fields), $\sigma$, and $\sigma^{\dagger}$
into  operators $\cO_i$\ that obey: 
\begin{equation}\label{eq:sigma-inv}
  \cO_i \to \mathbb{U} \cO_i \mathbb{U}^{\dagger}.
\end{equation}
We then construct chiral (and heavy-quark spin) invariants using these spurions
and the heavy-light fields.  At $\cO(a^2)$, exactly two taste spurions will appear in each
term in the Lagrangian and the current.
 In $\cL_{3,a^2}^{A}$ and $\cL_{3,a^2}^{B}$, which are
$\cO(ka^2)$, a derivative
term will also be needed. 
Either $D_\mu$ or $\mathbb{A}_\mu$ can be used.

A large class of contributions to 
$\cL^A_{2,a^2}$, $\cL^A_{3,a^2}$, and
$j^{\mu,b}_{2,a^2,A}$ can be obtained by first combining both
taste spurions into a single operator that transforms like \eq{sigma-inv}.
As described in \secref{NLOa2A}, all
such operators may be easily found from known results using a trick.
These operators are the same order in our chiral power
counting and have the same transformation properties 
as the quark mass terms, $\cM^\pm$. 
Thus, we can simply replace the $\cM^\pm$ in \eqsthree{L2m}{L3m}{j2m} 
with such two-taste-spurion  operators to find all terms in   
$\cL^A_{2,a^2}$, $\cL^A_{3,a^2}$, and
$j^{\mu,b}_{2,a^2,A}$ in which the two spurions combine together into a single
operator.  
It is then not hard to catalog the
remaining contributions, which are ones in which the two spurions are interspersed
between $\overline{H}H$ and $\mathbb{A}$ in a single trace, or are separately traced
with  $\overline{H}H$ or $\mathbb{A}$.

Determining the chiral representatives of 
$\cL^{\rm Sk}_B$ works similarly.
However each chiral invariant constructed from the spurions will
generate several terms in
$\cL^B_{2,a^2}$, $\cL^B_{3,a^2}$,
and $j^{\mu,b}_{2,a^2,B}$, due to various
ways of introducing the appropriate Lorentz indices using $v_\mu$, $\gamma_\mu$, $D_\mu$,
or $\mathbb{A_\mu}$.

\subsection{Discretization errors at NLO:  Operators from $\cL^{\rm Sk}_{A}$}
\label{sec:NLOa2A}

To find combinations of two taste spurions
from  $\cL^{\rm Sk}_{A}$ that transform like
\eq{sigma-inv}, we can start with 
the invariant light-meson operators that
result from the spurion analysis.
These are the eight operators given 
in \eq{VSigma} \cite{SCHPT,LEE_SHARPE}. 
From these operators, we can generate all ones that transform
like \eq{sigma-inv} by  simply replacing a factor of $\Sigma$  or $\Sigma^\dagger$
with, respectively, $\sigma^2$ or $(\sigma^\dagger)^2$, permuting the trace so there is 
a factor of $\sigma$ or $\sigma^\dagger$ at both beginning and end of the term, and
removing the trace.  
For example,
\begin{equation}\label{eq:opening}
\Tr(\xi^{(n)}_5\Sigma^{\dagger}\xi^{(n)}_5\Sigma) \to 
\sigma\xi^{(n)}_5\Sigma^{\dagger}\xi^{(n)}_5\sigma = \sigma\xi^{(n)}_5\sigma^{\dagger} \sigma^\dagger \xi^{(n)}_5\sigma\ ,
\end{equation}
where the final form shows explicitly how the result is composed of the produce of two taste spurions, each
transforming like \eq{sigma-inv}.
We call this process ``opening up'' a $\Sigma$ (or $\Sigma^\dagger$).
It automatically creates an operator that has the desired transformation
property when the spurions transform as required.  Each operator in \eq{VSigma} thereby
generates two operators, depending on whether one opens up a $\Sigma$ or 
a $\Sigma^\dagger$. As was done for the mass term in \eq{Mpm}, the two 
operators can be joined to make Hermitian or anti-Hermitian
combinations, or, equivalently, even or odd parity combinations.
This gives us the following operators that transform under chiral rotations by \eq{sigma-inv}:
\begin{eqnarray}\label{eq:OA}
  \cO^{A,\pm}_1 & = &  (\sigma\xi^{(n)}_5\Sigma^{\dagger}\xi^{(n)}_5\sigma 
  \pm p.c.)\nonumber \\*
  \cO^{A,\pm}_2 & = & \sum_{\nu}
  \left[ (\sigma\xi^{(n)}_{\nu}\sigma)\Tr(\xi^{(n)}_\nu\Sigma)
    \pm p.c.\right]\nonumber \\*
  \cO^{A,\pm}_3 & = & \sum_{\nu}( \sigma\xi^{(n)}_{\nu}\Sigma
    \xi^{(n)}_\nu\sigma \pm p.c.) \nonumber \\*
  \cO^{A,\pm}_4 & = & \sum_{\nu}( \sigma\xi^{(n)}_{\nu 5}\Sigma
    \xi^{(n)}_{5\nu}\sigma \pm p.c.) \nonumber \\*
  \cO^{A,\pm}_5 & = & \sum_{\nu}
  \left[ (\sigma\xi^{(n)}_{\nu}\sigma)\Tr(\xi^{(n)}_\nu\Sigma^\dagger)
  \pm p.c. \right]\nonumber \\*
  \cO^{A,\pm}_6 & = & \sum_\mu \sum_{\nu\not=\mu} 
  (\sigma\xi^{(n)}_{\mu\nu}\Sigma^{\dagger}
  \xi^{(n)}_{\nu\mu}\sigma \pm p.c.)  \nonumber \\*
  \cO^{A,\pm}_7 & = & \sum_{\nu}
  \left[ (\sigma \xi^{(n)}_{\nu5}\sigma)\Tr( \xi^{(n)}_{5\nu}\Sigma)
    \pm p.c.\right]\nonumber\\*
  \cO^{A,\pm}_8 & = & \sum_{\nu}
  \left[ (\sigma \xi^{(n)}_{\nu5}\sigma)\Tr(\xi^{(n)}_{5\nu}\Sigma^\dagger)
   \pm p.c. \right]\ ,
\end{eqnarray}
where $p.c.$ stands for parity conjugate: for example $\sigma_{p.c.}=\sigma^\dagger$.
We have followed the numbering scheme in \eq{VSigma}, with the exceptions that
the operators multiplied by $C_{2V}$ and $C_{5V}$ in \eq{VSigma}
become $\cO^{A,\pm}_2 $ and $\cO^{A,\pm}_5$, while those 
multiplied by $C_{2A}$ and $C_{5A}$ become $\cO^{A,\pm}_7 $ and $\cO^{A,\pm}_8$.
We have also not bothered to keep the conventional factors of $1/2$ and $1/4$ from \eq{VSigma}
in our definitions here.
Operators in \eq{OA} with no traces (coming from ones
in \eq{VSigma} with one trace) correspond to multiplying two taste spurions transforming like
\eq{sigma-inv}; 
operators with a trace (coming from ones
in \eq{VSigma} with two traces) correspond to tracing one taste spurion and then multiplying
by the other.

Making the replacement $\cM^\pm \to \cO^{A,\pm}_k$ in \eq{L2m}, we then get
\begin{equation}
  \cL^{A}_{2,a^2}  =  a^2\sum_{k=1}^8
  \Biggl\{K^A_{1,k} \Tr\left(\overline{H} H
   \cO^{A,+}_k\right)
  +  K^A_{2,k}\Tr\left(\overline{H} H\right)
	\Tr(  \cO^{A,+}_k) \Biggr\}\label{eq:L2A}
\end{equation}

$\cL^{A}_{3,a^2}$ is more complicated, since we can use one factor of $\mathbb{A}^\mu$. 
Because  $\mathbb{A}^\mu$ transforms like $\overline{H}H$, there is 
now an additional possibility of interspersing single  spurions 
between $\mathbb{A}^\mu$ and  $\overline{H}H$. A complete
spurion analysis is therefore required for such terms. We can follow
Appendix B of Ref.~\cite{LEE_SHARPE}, with the simple modifications \cite{SCHPT} that we
keep only ``odd-odd'' operators and make the spurions diagonal in flavor, as in \eq{xi-n}. 
We then prepend and append appropriate factors of $\sigma$ and $\sigma^\dagger$ 
to each spurion to make it into an operator that transforms like
\eq{sigma-inv}.  The result
of this analysis is that a given chiral operator of this kind in $\cL^{A}_{3,a^2}$ will involve a pair of
spurion operators, which we call
$Q^A_k$ and ${\tilde Q^A_k}$ ($k=1,\dots,8$):
\begin{eqnarray}\label{eq:QA}
  \big[V\times P\big],\; \big[A\times P\big]  \to \quad & Q^A_1  =   \sigma\xi^{(n)}_5\sigma^\dagger\;,&\quad {\tilde Q^A_1} \equiv 
  (Q^A_1)_{p.c.}=\sigma^\dagger\xi^{(n)}_5\sigma \nonumber \\*
  \big[V\times P\big],\; \big[A\times P\big]  \to \quad & Q^A_2  =   \sigma\xi^{(n)}_5\sigma^\dagger\;,&\quad {\tilde Q^A_2} \equiv 
  Q^A_2 \nonumber \\*
  \big[S\times V\big],\; \big[P\times V\big],\; \big[T\times V\big] \to \quad & Q^A_3  =   \sigma\xi^{(n)}_\nu\sigma\;,&\quad {\tilde Q^A_3} \equiv 
  Q^A_3 \nonumber \\*
  \big[S\times A\big],\; \big[P\times A\big],\; \big[T\times A\big] \ \to \quad & Q^A_4  =   i\sigma\xi^{(n)}_{\nu5}\sigma\;,&\quad {\tilde Q^A_4} \equiv 
  Q^A_4 \nonumber \\*
  \big[S\times V\big],\; \big[P\times V\big] \to \quad & Q^A_5  =   \sigma\xi^{(n)}_\nu\sigma\;,&\quad {\tilde Q^A_5} \equiv 
  (Q^A_5)_{p.c.}   =   \sigma^\dagger\xi^{(n)}_\nu\sigma^\dagger \nonumber \\*
  \big[V\times T\big],\; \big[A\times T\big]  \to \quad & Q^A_6  =   i\sigma\xi^{(n)}_{\lambda\nu}\sigma^\dagger\;,&\quad {\tilde Q^A_6} \equiv 
  (Q^A_6)_{p.c.}=-i\sigma^\dagger\xi^{(n)}_{\nu\lambda}\sigma \nonumber \\*
  \big[V\times T\big],\; \big[A\times T\big]  \to \quad & Q^A_7  =   i\sigma\xi^{(n)}_{\lambda\nu}\sigma^\dagger\;,&\quad {\tilde Q^A_7} \equiv 
  Q^A_7 \nonumber \\*
  \big[S\times A\big],\; \big[P\times A\big] \to \quad & Q^A_8  =   i\sigma\xi^{(n)}_{\nu5}\sigma\;,&\quad {\tilde Q^A_8} \equiv 
  (Q^A_8)_{p.c.}=-i\sigma^\dagger\xi^{(n)}_{5\nu}\sigma^\dagger\;, 
\end{eqnarray}
where we show the terms in $\cL^{\rm Sk}_A$ [\eq{SFFA}] that generate the given spurion operators.
The terms $[V\times S]$ and $[A\times S]$ in the Symanzik action 
do not appear in this list since the corresponding
spurion operators, such as $\sigma \xi^{(n)}_I \sigma^\dagger$, are trivial.
We have tried to make the numbering system in \eq{QA} correspond to that in
\eqs{VSigma}{OA} as much as possible.  Note that $Q^A_1=Q^A_2$  but ${\tilde Q^A_1}\not={\tilde Q^A_2}$,
and similarly for $Q^A_3=Q^A_5$, $Q^A_4=Q^A_8$, and $Q^A_6=Q^A_7$. Operators obtained by parity conjugating
both $Q^A_k$ and ${\tilde Q^A_k}$ are not included since they will appear automatically when we demand
parity invariance of the chiral Lagrangian.

We can now construct $\cL^{A}_{3,a^2}$ by first replacing $\cM^{\pm}\to \cO^{A,\pm}_k$
in \eq{L3m}, and then adding on new operators that intersperse $Q^A_k$ and ${\tilde Q^A_k}$
between $\overline{H}H$ and $\mathbb{A}^\mu$.
The result is:
\begin{eqnarray}
 	\cL^{A}_{3,a^2} & = &   a^2\sum_{k=1}^8\Biggl\{
    ic^A_{1,k} \Tr\left( \overline{H}H  v\negcdot \leftvec D\, \cO^{A,+}_k - v\negcdot \rightvec D \,
\overline{H}H\, \cO^{A,+}_k
  \right) \nonumber \\
&&\hspace{1.2truecm}+ i c^A_{2,k} 
 \Tr\left( \overline{H} H  v\negcdot \leftvec D -v\negcdot \rightvec D \,
\overline{H}H\right)
  \Tr(\cO^{A,+}_k) \nonumber\\
  &&\hspace{1.2truecm}+ 
    c^A_{3,k} \Tr\left( \overline{H}
     H \gamma_\mu\gamma_5\{\mathbb{A}^\mu,
    \cO^{A,+}_k\}\right)
  + c^A_{4,k} \Tr\left( \overline{H}
     H \gamma_\mu\gamma_5\mathbb{A}^\mu   \right)
    \Tr(\cO^{A,+}_k)
   \nonumber\\&&\hspace{1.2truecm}+
   c^A_{5,k}\Tr\left( \overline{H}
     H \gamma_\mu\gamma_5\right)\Tr(\mathbb{A}^\mu
    \cO^{A,+}_k)+ 
       c^A_{6,k} \Tr\left( \overline{H}
     H \gamma_\mu[\mathbb{A}^\mu,
    \cO^{A,-}_k]\right)\label{eq:L3A} \nonumber \\
&&\hspace{1.2truecm}+
c^A_{7,k} \Big( \Tr\big(\overline{H}
     H \gamma_\mu\gamma_5 Q^A_k \mathbb{A}^\mu
    {\tilde Q^A_k}\big) + p.c. \Big)
   \nonumber\\
&&\hspace{1.2truecm}+
c^A_{8,k} \Big( \Tr\big(\overline{H}
     H \gamma_\mu\gamma_5 Q^A_k\big)\Tr\big( \mathbb{A}^\mu
    {\tilde Q^A_k}\big) + p.c. \Big) \Biggr\}
   \nonumber\\
&&+a^2\sum_{k=2,5,7,8}
c^A_{9,k} \Big( \Tr\big(\overline{H}
     H \gamma_\mu Q^A_k\mathbb{A}^\mu {\tilde Q^A_k}\big) +p.c.\Big)\nonumber \\
&&+a^2\sum_{k=1,2,6,7}
c^A_{10,k} \Big( \Tr\big(\overline{H}
     H \gamma_\mu Q^A_k\big)\Tr\big( \mathbb{A}^\mu {\tilde Q^A_k}\big)+p.c.\Big) \ .
\end{eqnarray}
Here, any taste indices in $Q^A_k$ and $\tilde Q^A_k$ should be contracted; for example
one should sum over $\lambda$ and $\nu$ in a term involving  
$Q^A_6$, $\tilde Q^A_6$.\footnote{Contributions with $\lambda=\nu$ are automatically
omitted since we use the definition $\xi_{\lambda\nu}=(1/2)[\xi_\lambda,\xi_\nu]$.}
Note that only a subset of the $Q^A_k$ are allowed in the terms with coefficients
$c^A_{9,k}$ and $c^A_{10,k}$, which involve $\gamma_\mu$ as opposed to
$\gamma_\mu\gamma_5$. 
The parity conjugation introduces a minus sign, 
so the terms can only be Hermitian if $\tilde Q^A_k{}^\dagger= Q^A_k$ ($c^A_{9,k}$  term)
or ${Q^A_k}^\dagger=Q^A_k$ ($c^A_{10,k}$ term).
The forbidden terms are anti-Hermitian; adding a factor of $i$
to make them Hermitian would violate $\cT$ invariance.

The reader may wonder whether additional operators in $\cL^A_{3,a^2}$ might
be constructed using various $Q^A_j$ and ${\tilde Q^A_j}$ in place
of the $\cO_k^{A,\pm}$. The answer is no, since the corresponding terms
would contain either the product $Q^A_j{\tilde Q^A_j}$ or $Q^A_j\Tr({\tilde Q^A_j})$ (along with appropriate 
parity-conjugated terms).
In these circumstances, it is easy to check from  \eq{QA} that one reproduces
the $\cO_k^{A,\pm}$. 

The converse question is why the $Q^A_j$, ${\tilde Q^A_j}$ cannot be found by simply
opening up a second $\Sigma$ or $\Sigma^\dagger$ in the 
$\cO_k^{A,\pm}$.  The reason is that certain allowed combinations of spurions
--- namely 
$Q^A_2$,${\tilde Q^A_2}$ and $Q^A_7,{\tilde Q^A_7}$ ---
have been eliminated from  $ \cV_\Sigma $ and hence from $\cO_k^{A,\pm}$ because
their products are trivial and their traces vanish.  Such combinations
can only be found by returning to the original spurion analysis.

For the current at this order we cannot use $\mathbb{A}^\mu$, so
the replacement
$\cM^{\pm}\to \cO^{A,\pm}_k$ in \eq{j2m} is all that is needed.
We have
\begin{eqnarray}\label{eq:j2A}
  j^{\mu,b}_{2,a^2,A} & = & a^2\sum_{k=1}^8
  \Biggl\{
  r^A_{1,k}\,
  \trD\left(\gamma^\mu (1-\gamma_5) H  \right)
  \cO^{A,+}_k \sigma^\dagger \lambda^{(b)}
  + r^A_{2,k}\,
  \trD\left(\gamma^\mu (1-\gamma_5) H\right) \sigma^\dagger
   \lambda^{(b)} 
  \Tr(\cO^{A,+}_k) \nonumber \\
&&+
r^A_{3,k}\,
\trD\left(\gamma^\mu (1-\gamma_5) H\right)  
\cO^{A,-}_k \sigma^\dagger \lambda^{(b)}
+ r^A_{4,k}\,
\trD\left(\gamma^\mu (1-\gamma_5) H\right) \sigma^\dagger
 \lambda^{(b)}
\Tr(\cO^{A,-}_k) 
  \Biggr\}.
\end{eqnarray}

\subsection{Discretization errors at NLO:  Operators from $\cL^{\rm Sk}_{B}$}
\label{sec:NLOa2B}
The terms which arise from breaking rotation symmetry at 
this order can be determined through a spurion analysis 
that follows  that of Sec.~A3 in Ref.~\cite{Sharpe:2004is}.  
The following pairs of spurions can appear:
\begin{eqnarray}\label{eq:QB}
  \big[V_\mu\times T_\mu\big],\; \big[A_\mu\times T_\mu\big]  \to \quad & Q^B_{\mu,1}  =   i\sigma\xi^{(n)}_{\mu\lambda}\sigma^\dagger\;,&\quad {\tilde Q^B_{\mu,1}} \equiv 
  (Q^B_{\mu,1})_{p.c.}=-i\sigma^\dagger\xi^{(n)}_{\lambda\mu}\sigma \nonumber \\*
  \big[T_\mu\times V_\mu\big], \to \quad & Q^B_{\mu,2}  =   \sigma\xi^{(n)}_\mu\sigma\;,&\quad {\tilde Q^B_{\mu,2}} \equiv 
  (Q^B_{\mu,2})_{p.c.}   =   \sigma^\dagger\xi^{(n)}_\mu\sigma^\dagger\nonumber \\*
  \big[T_\mu\times A_\mu\big], \to \quad & Q^B_{\mu,3}  =   i\sigma\xi^{(n)}_{\mu5}\sigma\;,&\quad {\tilde Q^B_{\mu,3}} \equiv 
  (Q^B_{\mu,3})_{p.c.}=-i\sigma^\dagger\xi^{(n)}_{5\mu}\sigma^\dagger\nonumber \\*
  \big[V_\mu\times T_\mu\big],\; \big[A_\mu\times T_\mu\big]  \to \quad & Q^B_{\mu,4}  =   i\sigma\xi^{(n)}_{\mu\lambda}\sigma^\dagger\;,&\quad {\tilde Q^B_{\mu,4}} \equiv 
  Q^B_{\mu,4} \;,
\end{eqnarray}
where we show the terms in $\cL^{\rm Sk}_B$ [\eq{SFFB}] that generate the given spurion operators.
The index $\mu$ is singled
out in the names $Q^B_{\mu,j}$ because it will be repeated 
4 times  in terms in $\cL^{B}_{3,a^2}$,  thereby breaking separate continuous rotations and
taste symmetries, and leaving only the lattice symmetry of joint $90^\circ$ rotations. We do not
include the index $\lambda$ in the names $Q^B_{\mu,1}$ and $Q^B_{\mu,4}$ because $\lambda$ will appear
only twice, and  a sum over $\lambda$ will be implied 
in any terms with $Q^B_{\mu,1}$, $\tilde Q^B_{\mu,1}$ 
or $Q^B_{\mu,4}$, $\tilde Q^B_{\mu,4}$.
The spurions $Q^B_{\mu,j}$ ($j=1,2,3,4$) are in fact the same as $Q^A_6$,  
$Q^A_5$, $Q^A_8$ and $Q^A_7$, respectively.

We can also construct operators, analogous to the $\cO^{A,\pm}_k$, out 
of either $Q^B_{\mu,j}\tilde Q^B_{\mu,j}
\pm p.c.$ or $Q^B_{\mu,j}\Tr(\tilde Q^B_{\mu,j})
\pm p.c.$.
These are
\begin{eqnarray}\label{eq:OB}
    \cO^{B,\pm}_{\mu, 1} & = & \sum_{\lambda\not=\mu}
    (\sigma\xi^{(n)}_{\mu\lambda}\Sigma^\dagger\xi^{(n)}_{\lambda\mu}\sigma) 
    \pm p.c.\ ,\nonumber \\ 
    \cO^{B,\pm}_{\mu, 2} & = &  
    (\sigma\xi^{(n)}_\mu\sigma)
    \Tr(\xi^{(n)}_\mu\Sigma^\dagger) \pm p.c. \ ,\nonumber \\
    \cO^{B,\pm}_{\mu, 3} & = &  
    (\sigma\xi^{(n)}_{\mu5}\sigma)
    \Tr(\xi^{(n)}_{5\mu}\Sigma^\dagger) \pm p.c. \ .
\end{eqnarray}
The numbering here corresponds to that in \eq{QB}; $Q^B_{\mu,4}$, $\tilde Q^B_{\mu,4}$ produce
no operators of this kind.
Note that the repeated index $\mu$ is not summed over in \eq{OB}; aside from this these
operators are the same as operators $ \cO^{A,\pm}_6$,  $ \cO^{A,\pm}_5$, and $ \cO^{A,\pm}_8$, respectively. 
Like the $\cO^{A,\pm}_k$, the $\cO^{B,\pm}_{\mu,k}$  transform according to \eq{sigma-inv}.

To obtain terms that correspond to $\cL^{\rm Sk}_B$, we can multiply either the 
$ \cO^{B,\pm}_{\mu, k}$ or the $Q^B_{\mu,k}$, $\tilde Q^B_{\mu,k}$ 
by two additional four-vectors with index $\mu$,
and sum over $\mu$. For  $\cL^{B}_{2,a^2}$, we can employ $v_\mu$ or $\gamma_\mu$.
We find: 
\begin{equation}
	 \cL^{B}_{2,a^2} =  a^2\sum_\mu\sum_{k=1}^3
	 \Biggl\{ K^{B}_{1,k}v_\mu v^\mu \Tr(\overline{H} H
	 \cO^{B,+}_{\mu,k})  
	+ K^{B}_{2,k} v_\mu v^\mu  \Tr(\overline{H} H)
    \Tr(\cO^{B,+}_{\mu,k}) \Biggr\} \label{eq:L2B} 
\end{equation}
We have used that fact $\Tr(\overline{H}H\gamma_\mu)\propto \Tr(\overline{H}H)v_\mu$
to eliminate other possibilities.   Terms with $Q^B_{\mu,k}$, $\tilde Q^B_{\mu,k}$ 
are redundant and can be rewritten in terms of $ \cO^{B,\pm}_{\mu, k}$.

For $\cL^{B}_{3,a^2}$, a single factor of  
$D_\mu$ or $\mathbb{A}_\mu$ is also allowed. The enumeration then gets quite complicated,
because $Q^B_{\mu,k}$, $\tilde Q^B_{\mu,k}$  can appear in the
 $\mathbb{A}_\mu$ terms, and, in addition, various familiar simplifications cannot be used.
For example, manipulations that follow from $H \vslash = -H$ will not apply when the summation
index $\mu$ in $\vslash = v^\mu \gamma_\mu $ also appears in the taste-violating operators.
To make the equations more manageable, we write,
\begin{equation}
	\cL^{B}_{3,a^2} = \cL^{B,O}_{3,a^2}  + \cL^{B,Q}_{3,a^2} \ . \label{eq:L3B}
\end{equation}
We then obtain:
\begin{eqnarray}
    \cL^{B,O}_{3,a^2} &=&  a^2\sum_\mu\sum_{k=1}^3
	 \Biggl\{
    ic^B_{1,k} \Tr\left( \overline{H}H  v^\mu \leftvec D_\mu\, \cO^{B,+}_{\mu,k} - v^\mu \rightvec D_\mu \,
\overline{H}H\, \cO^{B,+}_{\mu,k}
  \right) \nonumber \\
&&\hspace{0.7truecm}+ i c^B_{2,k} 
 \Tr\left( \overline{H} H  v^\mu \leftvec D_\mu -v^\mu \rightvec D_\mu \,
\overline{H}H\right)
  \Tr(\cO^{B,+}_{\mu,k}) \nonumber\\
  &&\hspace{0.7truecm}+ 
    c^B_{3,k} \Tr\left( \overline{H}
     H \gamma_\mu\gamma_5\{\mathbb{A}^\mu,
    \cO^{B,+}_{\mu,k}\}\right)
  + c^B_{4,k} \Tr\left( \overline{H}
     H \gamma_\mu\gamma_5\mathbb{A}^\mu   \right)
    \Tr(\cO^{B,+}_{\mu,k})
   \nonumber\\&&\hspace{0.7truecm}+
   c^B_{5,k}\Tr\left( \overline{H}
     H \gamma_\mu\gamma_5\right)\Tr(\mathbb{A}^\mu
    \cO^{B,+}_{\mu,k})+ 
       c^B_{6,k} \Tr\left( \overline{H}
     H \gamma_\mu[\mathbb{A}^\mu,
    \cO^{B,-}_{\mu,k}]\right)\nonumber \\
    &&\hspace{0.7truecm}+ ic^B_{7,k} \, v_\mu v^\mu \Tr\left( \overline{H}H  v\negcdot \leftvec D\, \cO^{B,+}_{\mu,k} - v\negcdot \rightvec D \,
\overline{H}H\, \cO^{B,+}_{\mu,k}
  \right) \nonumber \\
&&\hspace{0.7truecm}+ i c^B_{8,k}\, v_\mu v^\mu  
 \Tr\left( \overline{H} H  v\negcdot \leftvec D -v\negcdot \rightvec D \,
\overline{H}H\right)
  \Tr(\cO^{B,+}_{\mu,k}) \nonumber\\
  &&\hspace{0.7truecm}+ 
    c^B_{9,k} \, v_\mu v^\mu \Tr\left( \overline{H}
     H \gamma_\nu\gamma_5\{\mathbb{A}^\nu,
    \cO^{B,+}_{\mu,k}\}\right)
  + c^B_{10,k} \, v_\mu v^\mu \Tr\left( \overline{H}
     H \gamma_\nu\gamma_5\mathbb{A}^\nu   \right)
    \Tr(\cO^{B,+}_{\mu,k})
   \nonumber\\&&\hspace{0.7truecm}+
   c^B_{11,k}\, v_\mu v^\mu \Tr\left( \overline{H}
     H \gamma_\nu\gamma_5\right)\Tr(\mathbb{A}^\nu
    \cO^{B,+}_{\mu,k})+ 
       c^B_{12,k} \, v_\mu v^\mu \Tr\left( \overline{H}
     H \gamma_\nu[\mathbb{A}^\nu,
    \cO^{B,-}_{\mu,k}]\right)\nonumber \\
  &&\hspace{0.7truecm}+ 
    c^B_{13,k}\, v^\mu \Tr\left( \overline{H}
     H \gamma_\mu\gamma_5\{v\negcdot \mathbb{A},
    \cO^{B,+}_{\mu,k}\}\right)
  + c^B_{14,k}\, v^\mu \Tr\left( \overline{H}
     H \gamma_\mu\gamma_5\, v\negcdot \mathbb{A}   \right)
    \Tr(\cO^{B,+}_{\mu,k})
   \nonumber\\&&\hspace{0.7truecm}+
   c^B_{15,k}\, v^\mu \Tr\left( \overline{H}
     H \gamma_\mu\gamma_5\right)\Tr(v\negcdot \mathbb{A}\,
    \cO^{B,+}_{\mu,k})
   +
    c^B_{16,k} v^\mu \Tr\left( \overline{H}
     H \gamma_{\mu\nu}\gamma_5\{\mathbb{A}^\nu,
    \cO^{B,+}_{\mu,k}\}\right)
   \nonumber\\&&\hspace{0.7truecm}+
    c^B_{17,k} v^\mu \Tr\left( \overline{H}
     H \gamma_{\mu\nu}\gamma_5\right)\Tr\left(\mathbb{A}^\nu
    \cO^{B,+}_{\mu,k}\right) 
+
    c^B_{18,k} v^\mu \Tr\left( \overline{H}
     H \gamma_{\mu\nu}\gamma_5\mathbb{A}^\nu \right)
    \Tr\left(\cO^{B,+}_{\mu,k}\right) 
   \nonumber\\&&\hspace{0.7truecm}+
    c^B_{19,k} v^\mu \Tr\left( \overline{H}
     H \gamma_{\mu\nu}\{\mathbb{A}^\nu,
    \cO^{B,-}_{\mu,k}\}\right)
+
    c^B_{20,k} v^\mu \Tr\left( \overline{H}
     H \gamma_{\mu\nu}\right)\Tr\left(\mathbb{A}^\nu
    \cO^{B,-}_{\mu,k}\right) 
    \Biggr\}
\label{eq:L3BO}
\end{eqnarray}
A sum over the repeated index $\nu$ is implied, and
 $\gamma_{\mu\nu}\equiv (1/2)[\gamma_\mu,\gamma_\nu]$.
We have used the fact that $\Tr( \cO^{B,-}_{\mu,k})$ vanishes for all $k$.

For  $\cL^{B,Q}_{3,a^2} $, we find:
\begin{eqnarray}
    \cL^{B,Q}_{3,a^2} &=&  a^2\sum_\mu
	 \Biggl\{
\sum_{k=1}^4 \bigg[
    c^B_{21,k} \Big(\Tr\big( \overline{H}
     H \gamma_\mu\gamma_5 Q^B_{\mu,k} \mathbb{A}^\mu
    \tilde Q^B_{\mu,k}\big) +p.c.\Big)
   \nonumber\\&&\hspace{2.3truecm}+
  c^B_{22,k} \Big(\Tr\big( \overline{H}
     H \gamma_\mu\gamma_5 Q^B_{\mu,k} \big) \Tr\big(\mathbb{A}^\mu \tilde Q^B_{\mu,k}  \big) +p.c.\Big)
   \nonumber\\&&\hspace{2.3truecm}+
    c^B_{23,k} v_\mu v^\mu \Big(\Tr\big( \overline{H}
     H \gamma_\nu\gamma_5 Q^B_{\mu,k} \mathbb{A}^\nu
    \tilde Q^B_{\mu,k}\big) +p.c.\Big)
   \nonumber\\&&\hspace{2.3truecm}+
  c^B_{24,k} v_\mu v^\mu \Big(\Tr\big( \overline{H}
     H \gamma_\nu\gamma_5 Q^B_{\mu,k} \big) \Tr\big(\mathbb{A}^\nu \tilde Q^B_{\mu,k}  \big) +p.c.\Big)
   \nonumber\\&&\hspace{2.3truecm}+
    c^B_{25,k} v^\mu \Big(\Tr\big( \overline{H}
     H \gamma_\mu\gamma_5 Q^B_{\mu,k} v\negcdot\mathbb{A}\,
    \tilde Q^B_{\mu,k}\big) +p.c.\Big)
   \nonumber\\&&\hspace{2.3truecm}+
  c^B_{26,k} v^\mu \Big(\Tr\big( \overline{H}
     H \gamma_\mu\gamma_5 Q^B_{\mu,k} \big) \Tr\big(v\negcdot\mathbb{A}\, \tilde Q^B_{\mu,k}  \big) +p.c.\Big)
   \nonumber\\&&\hspace{2.3truecm}+
    c^B_{27,k} v^\mu \Big( \Tr\big( \overline{H}
     H \gamma_{\mu\nu}\gamma_5 Q^B_{\mu,k} \mathbb{A}^\nu \tilde Q^B_{\mu,k} \big) +p.c.\Big)
   \nonumber\\&&\hspace{2.3truecm}+
    c^B_{28,k} v^\mu \Big( \Tr\big( \overline{H}
     H \gamma_{\mu\nu}\gamma_5 Q^B_{\mu,k}\big) \Tr\big( \mathbb{A}^\nu \tilde Q^B_{\mu,k} \big) +p.c.\Big)\bigg]
   \nonumber\\&&\hspace{0.8truecm}+
\sum_{k=2,3,4} \bigg[
    c^B_{29,k} \Big(\Tr\big( \overline{H}
     H \gamma_\mu Q^B_{\mu,k} \mathbb{A}^\mu
    \tilde Q^B_{\mu,k}\big) +p.c.\Big)
   \nonumber\\&&\hspace{2.3truecm}+
    c^B_{30,k} v_\mu v^\mu \Big(\Tr\big( \overline{H}
     H \gamma_\nu Q^B_{\mu,k} \mathbb{A}^\nu
    \tilde Q^B_{\mu,k}\big) +p.c.\Big)\bigg]
   \nonumber\\&&\hspace{0.8truecm}+
\sum_{k=1,4} \bigg[
  c^B_{31,k} \Big(\Tr\big( \overline{H}
     H \gamma_\mu Q^B_{\mu,k} \big) \Tr\big(\mathbb{A}^\mu \tilde Q^B_{\mu,k}  \big) +p.c.\Big)
   \nonumber\\&&\hspace{2.3truecm}+
  c^B_{32,k} v_\mu v^\mu \Big(\Tr\big( \overline{H}
     H \gamma_\nu Q^B_{\mu,k} \big) \Tr\big(\mathbb{A}^\nu \tilde Q^B_{\mu,k}  \big) +p.c.\Big) \bigg]
   \nonumber\\&&\hspace{0.8truecm}+
    c^B_{33,1} v^\mu \Big( \Tr\big( \overline{H}
     H \gamma_{\mu\nu} Q^B_{\mu,1} \mathbb{A}^\nu \tilde Q^B_{\mu,1} \big) +p.c.\Big)
   \nonumber\\&&\hspace{0.8truecm}+
\sum_{k=2,3} \bigg[
    c^B_{34,k} v^\mu \Big( \Tr\big( \overline{H}
     H \gamma_{\mu\nu} Q^B_{\mu,k}\big) \Tr\big( \mathbb{A}^\nu \tilde Q^B_{\mu,k} \big) +p.c.\Big)\bigg]
    \Biggr\}
\label{eq:L3BQ}
\end{eqnarray}
The additional implicit taste index $\lambda$ that appears in 
$Q^B_{\mu,1}$, $\tilde Q^B_{\mu,1}$ or
$Q^B_{\mu,4}$, $\tilde Q^B_{\mu,4}$ 
should be summed.
As in \eq{L3A},
only  a subset of the $Q^B_{\mu,k}$ are allowed by Hermiticity and $\cT$ invariance
in terms where the parity conjugate introduces a minus sign.  Since $\mathbb{A}$ is an axial vector,
these are the terms without an explicit $\gamma_5$.

Note that we have raised one space-time $\mu$ index in each term in \eqs{L3BO}{L3BQ}
because we are
using Minkowski space conventions.  We also emphasize that the $\gamma_\mu$ 
here, and elsewhere in the chiral Lagrangian, are always Minkowski ones.  The taste $\mu$ index on $\cO^{B,\pm}_{\mu,k}$
and $Q^B_{\mu,k}$, $\tilde Q^B_{\mu,k}$
is not affected by going to Minkowski space, and taste matrices $\xi_\mu$ obey Euclidean conventions.

Proceeding to the current, we have:
\begin{eqnarray}\label{eq:j2B}
  j^{\mu,b}_{2,a^2,B} & = & a^2\sum_{k=1}^3 \Biggl\{
  r^B_{1,k}\,
  \trD\left(\gamma^\mu (1-\gamma_5) H  \right)
  \cO^{B,+}_{\mu,k} \sigma^\dagger \lambda^{(b)}
  + r^B_{2,k}\,
  \trD\left(\gamma^\mu (1-\gamma_5) H\right) \sigma^\dagger
   \lambda^{(b)} 
  \Tr(\cO^{B,+}_{\mu,k}) \nonumber \\
&&\qquad+
r^B_{3,k}\,
\trD\left(\gamma^\mu (1-\gamma_5) H\right)  
\cO^{B,-}_{\mu,k} \sigma^\dagger \lambda^{(b)}
+ r^B_{4,k}\,
\trD\left(\gamma^\mu (1-\gamma_5) H\right) \sigma^\dagger
 \lambda^{(b)}
\Tr(\cO^{B,-}_{\mu,k}) \nonumber \\
  &&\qquad +\sum_\nu \Big( r^B_{5,k}\,
  \trD\left(\gamma^\mu (1-\gamma_5) H  \right)
  v_\nu v^\nu \cO^{B,+}_{\nu,k} \sigma^\dagger \lambda^{(b)}\nonumber \\
  &&\hspace{2.2truecm} + r^B_{6,k}\,
  \trD\left(\gamma^\mu (1-\gamma_5) H\right) \sigma^\dagger
   \lambda^{(b)} 
  v_\nu v^\nu \Tr(\cO^{B,+}_{\nu,k}) \nonumber \\
&&\hspace{2.2truecm} +
r^B_{7,k}\,
\trD\left(\gamma^\mu (1-\gamma_5) H\right)  
v_\nu v^\nu \cO^{B,-}_{\nu,k} \sigma^\dagger \lambda^{(b)}\nonumber \\
&&\hspace{2.2truecm} + r^B_{8,k}\,
\trD\left(\gamma^\mu (1-\gamma_5) H\right) \sigma^\dagger
 \lambda^{(b)}
v_\nu v^\nu \Tr(\cO^{B,-}_{\nu,k}) \nonumber \\
  &&\hspace{2.2truecm}  +r^B_{9,k}\,
  \trD\left(\gamma^\mu (1-\gamma_5) H \gamma^\nu  \right)
  v_\nu  \cO^{B,+}_{\nu,k} \sigma^\dagger \lambda^{(b)}\nonumber \\
  &&\hspace{2.2truecm} + r^B_{10,k}\,
  \trD\left(\gamma^\mu (1-\gamma_5) H \gamma^\nu\right) \sigma^\dagger
   \lambda^{(b)} 
  v_\nu  \Tr(\cO^{B,+}_{\nu,k}) \nonumber \\
&&\hspace{2.2truecm} +
r^B_{11,k}\,
\trD\left(\gamma^\mu (1-\gamma_5) H \gamma^\nu\right)  
v_\nu  \cO^{B,-}_{\nu,k} \sigma^\dagger \lambda^{(b)}\nonumber \\
&&\hspace{2.2truecm} + r^B_{12,k}\,
\trD\left(\gamma^\mu (1-\gamma_5) H \gamma^\nu\right) \sigma^\dagger
 \lambda^{(b)}
v_\nu  \Tr(\cO^{B,-}_{\nu,k}) 
  \Big)\Biggr\}\ .
\end{eqnarray}
Here the repeated index $\mu$ is not summed over.

Despite the complexity of \eqsthru{L2A}{j2B}, the effect of these terms on 
the heavy-light decay constant is extremely simple.
At this order, the terms can contribute only at tree level, so we can set $\sigma=1$ and $\mathbb{A_\mu}=0$. 
Each term then either vanishes outright, is nonvanishing but does not contribute to $f_B$, or 
reduces to an additive constant in $f_B$ that is proportional to $a^2$.
In particular, there are no rotation-violating contributions to $f_B$ until higher order.
All these $a^2$ analytic terms thus combine and lead merely to the presence of
one additional unknown parameter  in \schpt\ chiral fits, 
compared to continuum-like chiral fits.

\section{Chiral Logarithms in $f_B$ at one loop}
\label{sec:fB}

Before calculating the one-loop diagrams, we need to write down the propagators of the
heavy-light and pion fields, as well as the vertices coupling them.
We use the notation $\{\psi\chi\}(k)$ to denote the Minkowski space propagator of 
fields $\psi$ and $\chi$ with momentum $k$. 
Then, from the LO heavy-light Lagrangian $\cL_1$, \eq{L1}, we have 
\begin{eqnarray}
	\Bigl\{B_a B^\dagger_b\Bigr\}(k) &=& \frac{i\delta_{ab}}{2 (v\negcdot k + i\epsilon)}\ , \label{eq:Bprop}\\
	\Bigl\{B^*_{\mu a} B^{*\dagger}_{\nu b}\Bigr\}(k) &=& \frac{-i\delta_{ab}(g_{\mu\nu} - v_\mu v_\nu)}
	{2(v\negcdot k + i\epsilon)} \ \label{eq:Bstarprop}.
\end{eqnarray}
Here $a,b$ indicate the flavor-taste of the light quarks.
Lower case Latin indices from the beginning of the alphabet ($a,b,c,d\dots$)
will always serve as such flavor-taste indices; in the full QCD case they run over $4N_f$ values, where $N_f$
is the number of light sea-quark flavors. 

Since taste violations do not appear in $\cL_1$,
\eqs{Bprop}{Bstarprop} are the same as in the continuum theory \cite{MAN_WISE},
except that flavor-taste indices replace pure flavor indices.
Similarly, the $BB^*\pi$ vertex looks identical to that in Ref.~\cite{MAN_WISE} aside
from the redefinition of the indices and a factor of $1/2$ in the present case coming from
our normalization of the taste generators, \eq{T_Xi}.  The term in the interaction
Lagrangian that gives this vertex is:
\begin{equation}\label{eq:B-Bstar-pi}
\frac{ig_\pi}{f}\,\left(B^{*\dagger}_{\mu a}\, B_b \, -
B^\dagger_a \, B^{*}_{\mu b}\right) \, \partial^\mu \Phi_{ba} \ ,
\end{equation}
where repeated indices are summed.
The other needed vertex comes from the $B\pi\pi$ term in the expansion of
the LO current, \eq{LOcurrent}.  We have:
\begin{equation}\label{eq:current-vertex}
j^{\mu,b}_{\rm LO} = i\kappa v^\mu\left( B_b -\frac{1}{8f^2}\,B_a \, 
\Phi_{ac} \Phi_{cb} + \cdots \right) \ ,
\end{equation}
where  repeated indices are again summed and $\cdots$ represents terms involving $B^*$ or other numbers of pions, which do not
contribute to the decay constant at this order.

When necessary, each flavor-taste index can be replaced by a pair of indices representing flavor and taste separately.
We use Latin indices in the middle of the alphabet ($i,j,\dots$) as pure flavor indices, which take on the
values $1,2,\dots,N_f$ in full QCD.
Greek indices at the beginning of the alphabet ($\alpha,\beta,\gamma,\dots$) 
will be used for quark taste indices,
running from $1$ to $4$.  Thus we can replace $a\to i\alpha$ and write, for example,
\begin{equation}
	\Bigl\{B_{i\alpha} B^\dagger_{j\beta}\Bigr\}(k) = \frac{i\delta_{ij}\delta_{\alpha\beta}}{2 (v\negcdot k + i\epsilon)}\ .
\end{equation}

For light mesons, the LO Lagrangian $\cL_{\rm LO}^{\rm pions}$, \eq{LLO}, includes both the continuum terms
$\cL_{\rm LO,cont}^{\rm pions}$ and the leading $a^2$ corrections 
$\cL_{{\rm LO},a^2}^{\rm pions}$. As in Ref.~\cite{SCHPT}, we treat explicitly the
``hairpin'' terms arising from the $m_0^2$ contribution to  \eq{Lcontpion} and the
$C_{2V}$, $C_{2A}$, $C_{5V}$, and $C_{5A}$ contributions to \eq{VSigma} by separating
out the disconnected parts of meson propagators.  The remainder of $\cL_{\rm LO}^{\rm pions}$ 
determines the connected pion propagators:
\begin{equation}\label{eq:PropConnTaste}
	\Bigl\{\Phi^{\Xi}_{ij}\Phi^{\Xi'}_{j'i'}\Bigr\}_{\rm conn}(p) = 
	\frac{i\delta_{ii'}\delta_{jj'} \delta_{\Xi\Xi'}}{p^2 - m_{ij,\Xi}^2 + i\epsilon}
\end{equation}
where $\Xi$ is one of the 16 meson tastes [as defined after \eq{Phi}], and $m_{ij,\Xi}$ is
the tree-level mass of a taste-$\Xi$ meson composed of quarks of flavor $i$ and $j$:
\begin{equation}\label{eq:pi-masses-specific}
       m_{ij,\Xi}^2  = \mu (m_i + m_j) + a^2\Delta_\Xi.
\end{equation}
Here $\Delta_\Xi$ is the taste splitting, which can be expressed in terms of 
$C_1$, $C_3$, $C_4$ and $C_6$ in \eq{VSigma} \cite{SCHPT}.

Because of the residual $SO(4)$ taste
symmetry \cite{LEE_SHARPE} at this order, the mesons within a given
taste multiplet ($P$, $V$, $T$, $A$, or $I$) are degenerate in mass. 
When it is not important to specify
the particular member of the multiplet in question, we
 will usually just name the multiplet, for example:
\begin{equation}\label{eq:pi-masses}
       m_{ij,V}^2  = \mu (m_i + m_j) + a^2\Delta_V.
\end{equation}

Since the heavy-light propagators are most simply written with flavor-taste indices,
as in \eqs{Bprop}{Bstarprop}, it is convenient for current purposes to rewrite
\eq{PropConnTaste} in flavor-taste notation also:
\begin{equation}\label{eq:PropConn}
	\Bigl\{\Phi_{ab}\Phi_{b'a'}\Bigr\}_{\rm conn}(p) \equiv 
	\Bigl\{\Phi_{i\alpha, j\beta}\Phi_{j'\beta',i'\alpha'}\Bigr\}_{\rm conn}(p) = \sum_\Xi
	\frac{i\delta_{ii'}\delta_{jj'} T^\Xi_{\alpha\beta}  T^\Xi_{\beta'\alpha'} }
{p^2 - m_{ij,\Xi}^2 + i\epsilon} \ ,
\end{equation}
where $T^\Xi$ are the 16 taste generators, \eq{T_Xi}.

For flavor-charged pions ($i\not=j$), the complete propagators are just
the connected propagators in \eqsor{PropConnTaste}{PropConn}.  However, for flavor-neutral
pions ($i=j$), there are disconnected contributions coming from one or more
hairpin insertions.  At LO, these appear only for taste singlet, vector,
or axial-vector pions. 
Denoting the Minkowski hairpin
vertices as $-i\delta'_\Xi$, we have
\cite{SCHPT}:
\begin{equation}\label{eq:dp_def}
  \delta_\Xi' = \begin{cases}
    a^2 \delta'_V, &T_\Xi\in\{\xi_\mu\}\ \textrm{(taste\ vector);}\\*
    a^2 \delta'_A, &T_\Xi\in\{\xi_{\mu5}\}\ \textrm{(taste\ axial-vector);}\\*
    4m_0^2/3, &T_\Xi=\xi_{I}\ \textrm{(taste\ singlet;)}\\*
    0, &T_\Xi\in\{\xi_{\mu\nu},\xi_5\}\ \textrm{(taste\ tensor or pseudoscalar)}
  \end{cases}
\end{equation}
with
\begin{eqnarray}\label{eq:mix_vertex_VA}
 \delta'_{V(A)} & \equiv & \frac{16}{f^2} (C_{2V(A)} - C_{5V(A)})\ .
\end{eqnarray}
The disconnected pion propagator is then
\begin{equation}\label{eq:PropDiscTaste}
	\Bigl\{\Phi^{\Xi}_{ij}\Phi^{\Xi'}_{j'i'}\Bigr\}_{\rm disc}(p)=
	\delta_{ij}\delta_{j'i'} \delta_{\Xi\Xi'} \cD^\Xi_{ii,i'i'} \ ,
\end{equation}
where \cite{SCHPT}
 \begin{equation}\label{eq:Disc}
\cD^\Xi_{ii,i'i'} = -i\delta'_\Xi \frac{i}{(p^2-m_{ii,\Xi}^2)}
 \frac{i}{(p^2-m_{i'i',\Xi}^2)}
  \frac{(p^2-m_{U,\Xi}^2)(p^2-m_{D,\Xi}^2)(p^2-m_{S,\Xi}^2)}
       {(p^2-m_{\pi^0,\Xi}^2)(p^2-m_{\eta,\Xi}^2)(p^2-m_{\eta',\Xi}^2)}\ .
\end{equation}
For concreteness we have assumed that there are three sea-quark flavors: $u$, $d$, and $s$;
the generalization to $N_f$ flavors is immediate.  
Here $m_{U,\Xi}\equiv m_{uu,\Xi}$ is the mass of a taste-$\Xi$  pion made from a $u$ and a $\bar u$ quark,
neglecting hairpin mixing (and similarly for $m_{D,\Xi}$ and $m_{S,\Xi}$),  $m_{\pi^0,\Xi}$, $m_{\eta,\Xi}$, and $m_{\eta',\Xi}$
are the mass eigenvalues after mixing is included, and
the $i\epsilon$ terms have been left implicit.  When specifying the particular member
of a taste multiplet appearing in the disconnected
propagator is unnecessary, we will abuse this notation slightly following
\eq{pi-masses} and
refer to $\cD^V_{ii,i'i'} $, $\cD^A_{ii,i'i'}$, or $\cD^I_{ii,i'i'}$. 
In flavor-taste notation we have:
\begin{equation}\label{eq:PropDisc}
	\Bigl\{\Phi_{ab}\Phi_{b'a'}\Bigr\}_{\rm disc}(p) \equiv 
	\Bigl\{\Phi_{i\alpha, j\beta}\Phi_{j'\beta',i'\alpha'}\Bigr\}_{\rm disc}(p) = 
	\delta_{ij}\delta_{j'i'} 
\sum_\Xi
T^\Xi_{\alpha\beta}  T^\Xi_{\beta'\alpha'} \cD^\Xi_{ii,i'i'}
\end{equation}

\Eqsthree{PropDiscTaste}{Disc}{PropDisc} apply both to full and partially quenched QCD.
In the latter case, $i$ and $i'$ can represent 
(quenched) valence quarks. Below, we will need the disconnected propagator only
when $i$ and $i'$ both denote a  particular valence quark: call it flavor $x$. Letting
$X$ be the flavor-neutral meson consisting of an $x$ and an $\bar{x}$, we will use
the simplified notation $\cD^\Xi_{XX}\equiv \cD^\Xi_{xx,xx}$
or, for example, $\cD^V_{XX}\equiv \cD^V_{xx,xx}$.

We now calculate the one-loop decay constant, $f_{B_a}$, of a heavy-light pseudoscalar meson with
light flavor-taste $a$.  
The calculation is done in the \fpfpf\ 
theory, where there are four tastes per flavor. We then
adjust the result for the \opopo\ (one taste per flavor) case of interest.
We work out explicitly the partially quenched case;
full theory results are easily obtained
by taking appropriate limits. 
The valence flavor is called $x$; the valence taste, $\alpha$.  In other words, we replace $a \to x\alpha$.
As will be seen, the decay constant does not depend on the taste of the valence quark, so
we may write $f_{B_a} \to f_{B_x}$.

\begin{figure}
 \includegraphics[width=3in]{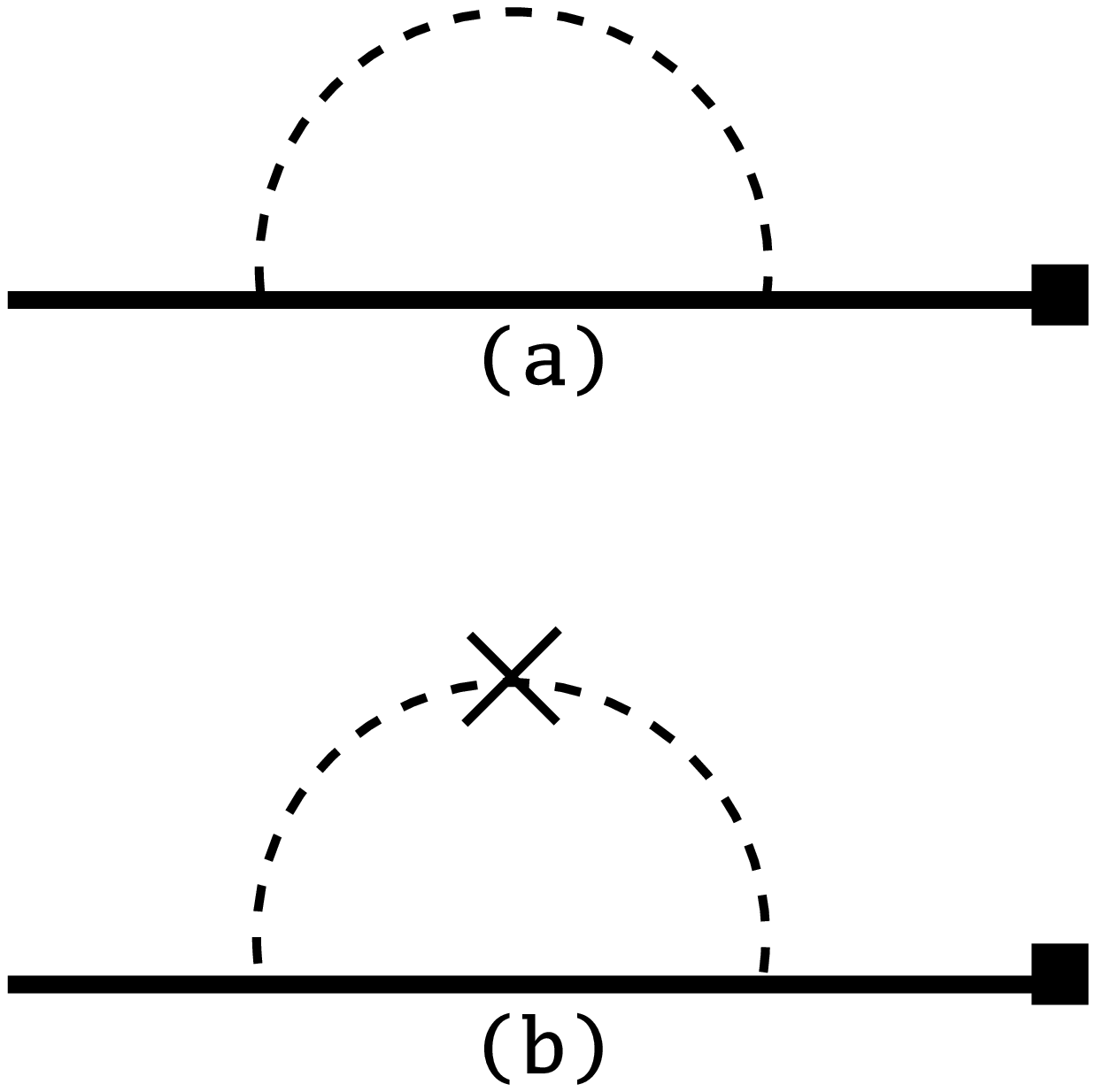} \caption{Terms which
 contribute to the one-loop heavy-light meson decay constant,
 arising from wavefunction renormalization. The
 thick solid line is the heavy-light meson, and the dashed line is the
 light pseudoscalar (\ie ``pion''). The solid square is the current insertion.
The cross indicates a disconnected propagator, \eq{PropDisc}; a pion line without a cross is connected.}
  	\label{fig:tadZ}
\end{figure}

\begin{figure}
 \includegraphics[width=3in]{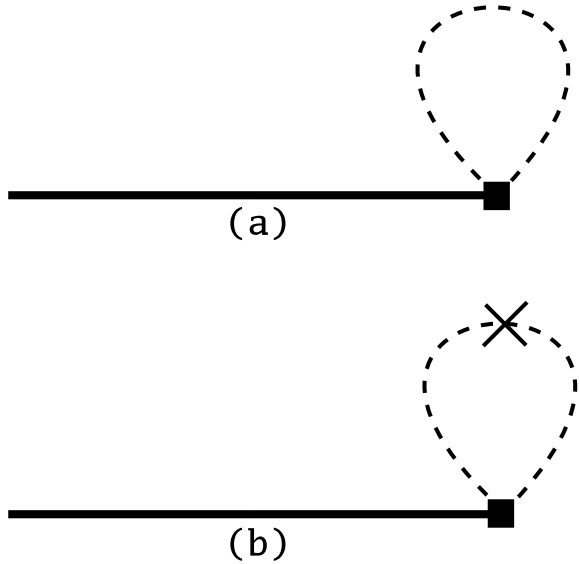} \caption{
 Contributions to the one-loop heavy-light meson decay constant
 coming from corrections to the current insertion. Symbols have
the same meaning as in 	Fig.~\ref{fig:tadZ}.  }
  	\label{fig:tadF}
\end{figure}

\begin{figure}
 \includegraphics[width=3in]{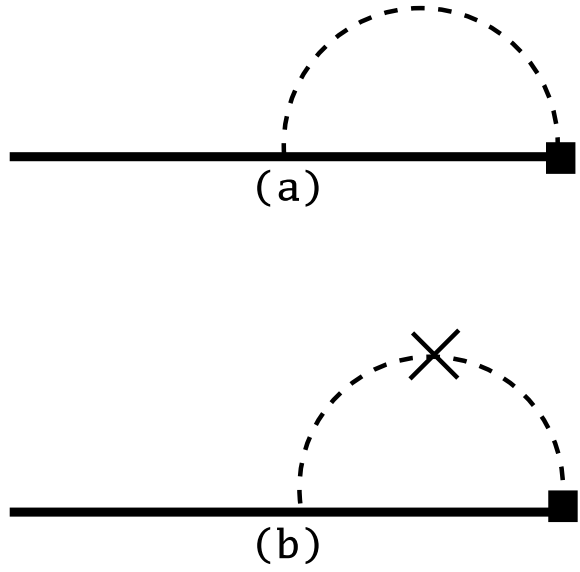} \caption{Terms which in
 principle arise at one loop in the decay constant but vanish
 since they are proportional to $v^\nu (g_{\mu\nu} - v_\mu v_\nu)$.}
  	\label{fig:tadZ_zero}
\end{figure}

At one loop, we express the decay constant as
\begin{equation}\label{eq:fB}
        f_{B_{x}} = f_{B_{x}}^{\rm LO} \left( 1 + \frac{1}{16\pi^2 f^2}\;
        \delta\! f_{B_{x}} + {\rm analytic\ terms} \right) \ .
\end{equation}
Recall that the lowest order term $f_{B_{x}}^{\rm LO}$ depends on the light valence
flavor in a rather trivial way: $f_{B_{x}}^{\rm LO} = \kappa/\sqrt{m_{B_{x}}}$.
The one-loop diagrams that contribute to $\delta\! f_{B_{x}}$  are
shown in Figs.~\ref{fig:tadZ} and \ref{fig:tadF}, where the cross
indicates a disconnected pion propagator (\ie one or more hairpin insertions for the singlet,
vector and axial tastes). The terms in
Fig.~\ref{fig:tadZ_zero}
vanish because the integrals produce an overall
factor of $v_\nu$ on shell, which is then multiplied by the projector $g^{\mu\nu} -
v^{\mu}v^{\nu}$.  

The diagrams in Fig.~\ref{fig:tadZ} contribute to the heavy-light wavefunction renormalization.
Using the propagators and vertices defined above, one easily sees that the
self energy $S_{ab}$ is proportional to the identity in taste space, and is diagonal
in flavor. We thus have
\begin{equation}
  S_{ab}(v\!\cdot\! k) \equiv  S_{i\alpha,j\beta}(v\!\cdot\! k) =  \delta_{\alpha\beta}\,
\delta_{ij}\, \tilde S_{i}(v\!\cdot\! k)  \ .
\end{equation}
Flavor symmetry (broken by the diagonal mass terms) guarantees this flavor structure. 
The taste-independence follows from 
the residual discrete taste symmetry of the light quarks,
which corresponds to the shift symmetry in the staggered action. 
This symmetry is
\begin{equation}
  q_i \to (1\otimes \xi_\mu) q_i \,,\quad
  \bar q_i \to \bar q_i (1 \otimes \xi_\mu) \ ,
  \label{eq:disc_shift}
\end{equation}
at the level of the Symanzik action; while at the chiral
level it is
\begin{eqnarray}\label{eq:shift_chiral}
  \Sigma & \to & \xi^{(n)}_\mu \Sigma \xi^{(n)}_\mu\ , \nonumber\\
  \sigma & \to & \xi^{(n)}_\mu \sigma \xi^{(n)}_\mu\ , \nonumber\\
  H    & \to & H \xi^{(n)}_\mu \ ,\nonumber\\
  \overline{H}    & \to & \xi^{(n)}_\mu \overline{H}\ .
\end{eqnarray}
Note that the symmetry
is diagonal in flavor; the transformation acts only on the taste
indices and affects all light quark flavors identically.

The one-loop wavefunction renormalization, $\delta Z_{B_x}$,  for $B_x$ is then
\begin{equation}
  \delta Z_{B_x} = \frac{1}{2}\frac{d \tilde S_x(v\!\cdot\! k)}{d(v\!\cdot\!
  k)}\Biggr\vert_{v\cdot k=0}\ .
\end{equation}
This contributes an amount $ \delta f_{B_{x}}^{Z}$ to $\delta f_{B_x}$. 
We find
(in the $\fpfpf$ theory):
\begin{eqnarray}\label{eq:initial-fB-Z}
  \delta f_{B_{x}}^{Z} & = & \frac{1}{2}\,( 16\pi^2f^2)\,  \delta 
  Z_{B_{x}} 
  \nonumber \\* 
  & = & \frac{g_\pi^2}{8}\left(
  g_{\mu\nu} - v_\mu v_\nu \right) \int
  \frac{d^4 p}{\pi^2} \frac{p^\mu p^\nu }{(v\cdot p+i\epsilon)^2}  
   \Biggl[\sum_{f,\Xi}\frac{i}{p^2-m^2_{xf,\Xi}+i\epsilon} \nonumber\\*
     &&\hspace{7.0truecm}{}+ \cD^I_{XX} + 4\cD^V_{XX} + 
     4\cD^A_{XX} \Biggr]
\end{eqnarray}
where $\Xi$ runs over the 16 pion tastes, and 
$f$ runs over the three sea quarks, $u$, $d$, and $s$.  The term summed over
$f$ and $\Xi$ arises from the ``connected'' diagram, \figref{tadZ}(a); as we will
see below, it involves an internal sea quark loop. 
The remaining terms come from the ``disconnected'' diagram,  \figref{tadZ}(b).
We have summed over the four degenerate vector and axial disconnected contributions, as
seen by the factors of four in front of $\cD^V$ and $\cD^A$.

The current corrections
contribute an amount $ \delta f_{B_{x}}^{\rm cur}$ to $\delta f_{B_x}$. Computing
the diagrams in \figref{tadF}, we obtain (in the $\fpfpf$ theory):
\begin{eqnarray}\label{eq:initial-fB-cur}
  \delta f_{B_{x}}^{\rm cur}   & = & -\frac{1}{8} \int
  \frac{d^4 p}{\pi^2}  
   \left[\sum_{f,\Xi}\frac{i}{p^2-m^2_{xf,\Xi}+i\epsilon} + \cD^I_{XX} + 4\cD^V_{XX} + 
     4\cD^A_{XX} \right]\ .
\end{eqnarray}
These chiral integrals are similar to those that appear 
\cite{SCHPT} for
the decay constant, $f_P$, of
a partially quenched pion $P$. 
The main difference is that here we have a single
light valence quark $x$, whereas $P$ is a bound state of two light valence quarks, $x$ and $\bar y$.
So the corresponding integrals in the $f_P$ case have additional 
contributions with $x\to y$ in various terms.

\begin{figure}
 \includegraphics[width=4in]{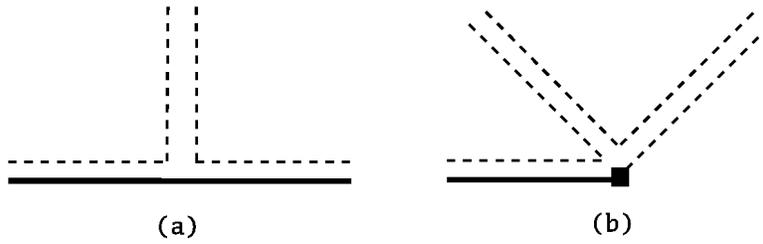} \caption{Relevant vertices
 at the quark level. (a) is the $g_\pi$ vertex. (b) is the
 second-order diagram coming from the current.}
  	\label{fig:vertices}
\end{figure}

\begin{figure}
 \includegraphics[width=3in]{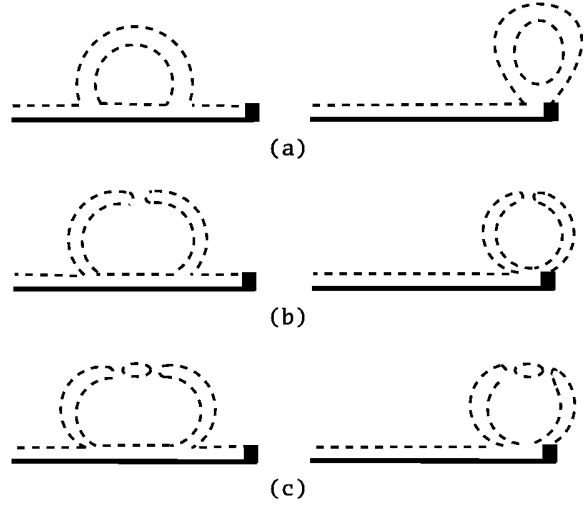} \caption{Quark-flow
   diagrams that contribute to $f_B$, 
   corresponding to Figs.~\ref{fig:tadZ} (left hand side) and \ref{fig:tadF}
   (right hand side).
   The dashed lines here are the light quarks while solid
   lines are the heavy quarks.}
  	\label{fig:quarkflow}
\end{figure}

We now need to convert \eqs{initial-fB-Z}{initial-fB-cur} from a $\fpfpf$ to a $\opopo$ theory.
This can easily be done with a quark-flow analysis, following Ref.~\cite{SCHPT}.
The $g_\pi$ vertex at the quark level is
shown in Fig.~\ref{fig:vertices}(a); 
the current vertex at second order in the
pion fields is depicted in Fig.~\ref{fig:vertices}(b). 
The connected terms in \eqs{initial-fB-Z}{initial-fB-cur}
that come from these vertices thus involve an internal quark
loop, shown  in Fig.~\ref{fig:quarkflow}(a), and should be multiplied
by a factor of $1/4$.
As described in Ref.~\cite{SCHPT}, the only
other changes in going from $\fpfpf$ to $\opopo$ appear in the mass eigenstates of the full
flavor-neutral propagators. This is due to the factors of $1/4$ that are associated with
iteration of the two-point vertex, \eq{dp_def}, in 
Fig.~\ref{fig:quarkflow}(c).  

In the $\opopo$ theory, we thus have
\begin{eqnarray}\label{eq:fB-Z}
  \delta f_{B_{x}}^{Z} 
  & = & \frac{g_\pi^2}{8}\left(
  g_{\mu\nu} - v_\mu v_\nu \right) \int
  \frac{d^4 p}{\pi^2} \frac{p^\mu p^\nu }{(v\cdot p+i\epsilon)^2}  
   \Biggl[\frac{1}{4}\sum_{f,\Xi}\frac{i}{p^2-m^2_{xf,\Xi}+i\epsilon} \nonumber\\*
     &&\hspace{7.0truecm}{}+ \cD^I_{XX} + 4\cD^V_{XX} + 
     4\cD^A_{XX} \Biggr] \\
\label{eq:fB-cur}
  \delta f_{B_{x}}^{\rm cur}   & = & -\frac{1}{8} \int
  \frac{d^4 p}{\pi^2}  
   \left[\frac{1}{4}\sum_{f,\Xi}\frac{i}{p^2-m^2_{xf,\Xi}+i\epsilon} + \cD^I_{XX} + 4\cD^V_{XX} + 
     4\cD^A_{XX} \right]\ ,
\end{eqnarray}
where the mass eigenvalues that appear in $\cD^I$, $\cD^V$,
and $\cD^A$ [\cf \eq{Disc}] are now the $\opopo$ values \cite{SCHPT}.  
For example, with three sea-quark flavors, $m_{\eta',I}\sim m_0^2$ for large $m_0^2$, rather 
than the value $4m_0^2$  that follows from \eq{Lcontpion}.  
After the $m_0^2\to\infty$ limit is taken,
this implies that the
$\cD^I$ contribution is 4 times larger in the  \opopo\ theory than in the $\fpfpf$ theory.

The disconnected propagators in \eqs{fB-Z}{fB-cur} can then be written as a sum of single
or double poles using the residue functions introduced in Ref.~\cite{SCHPT}.
We define $\{m\}\equiv \{m_1,m_2,\dots,m_n\}$ as the set of masses that appear in the
denominator of \eq{Disc}, and $\{\mu\}\equiv \{\mu_1,\mu_2,\dots,\mu_k\}$ as the numerator
set of masses.  Then, for $n>k$ and all masses distinct, we have:
\begin{equation}\label{eq:lagrange}
        \cI^{[n,k]}\left(\left\{m\right\}\!;\!\left\{\mu\right\}\right)
        \equiv \frac{\prod_{i=1}^k (q^2 - \mu^2_i)}
               {\prod_{j=1}^n (q^2 - m^2_j + i\epsilon)} =
        \sum_{j=1}^n \frac{(-1)^{n+k+1}R_j^{[n,k]}\left(\left\{m\right\}\!;\!
          \left\{\mu\right\}\right)}{q^2 - m^2_j + i\epsilon}\ ,
\end{equation}
where
\begin{equation}\label{eq:residues}
        R_j^{[n,k]}\left(\left\{m\right\}\!;\!\left\{\mu\right\}\right)
         \equiv  \frac{\prod_{i=1}^k (\mu^2_i- m^2_j)}
        {\prod_{r\not=j} (m^2_r - m^2_j)}\ .
\end{equation}
If there is one double pole term for $q^2=m_\ell^2$ (where $m_\ell \in \{m\}$), then
\begin{eqnarray}
        \cI^{[n,k]}_{\rm dp}\left(m_{\ell};\left\{m\right\}\!
        ;\!\left\{\mu\right\}\right) 
        &\equiv& \frac{\prod_{i=1}^k (q^2 - \mu^2_i)}
        {(q^2 - m^2_{\ell}+i\epsilon )\prod_{j=1}^{n} (q^2 - m^2_j+i\epsilon )}
        \nonumber\\*
        & = & \frac{\partial}{\partial m^2_\ell}   \sum_{j=1}^n 
        \frac{(-1)^{n+k+1}R_j^{[n,k]}\left(\left\{m\right\}\!;\!
          \left\{\mu\right\}\right)}{q^2 - m^2_j +i\epsilon}\label{eq:lagrange2}\ .
\end{eqnarray}
We use the same definitions of the residues $R^{[n,k]}_j$ as in
Ref.~\cite{SCHPT}; the factors of $(-1)^{n+k+1}$ arise from the fact
that the residues were defined for Euclidean space integrals, but we are 
working in Minkowski space here. 

Using \eqs{lagrange}{lagrange2},
the integration in \eq{fB-cur} is immediate, and that in \eq{fB-Z} is
standard in the continuum heavy-light literature.  We follow Ref.~\cite{BOYD} and define
(in $4-\hat\epsilon$ dimensions, with chiral scale $\Lambda_\chi$)
\begin{eqnarray}
  J^{\mu\nu}(m,\Delta) & \equiv & i\Lambda_\chi^{\hat\epsilon} \int \frac{d^{4-\hat\epsilon} p}
{(2\pi)^{4-\hat\epsilon}} \frac{p^\mu
    p^\nu }{(p^2-m^2+i\epsilon)(v\cdot p - \Delta+i\epsilon)}\nonumber \\
&=& \frac{1}{16\pi^2}\Delta[J_1(m,\Delta)g^{\mu\nu}+J_2(m,\Delta)v^\mu v^\nu] \ .\label{eq:Jmunu}
\end{eqnarray}
$\Delta$ in this equation should not be confused with the $\Delta_\Xi$ defined in \eq{pi-masses-specific}.
For \eq{fB-Z}, we need only 
\begin{eqnarray}
  (g_{\mu\nu} - v_\mu v_\nu)\frac{\partial J^{\mu\nu}(m,\Delta)}
  {\partial \Delta}\Biggr\vert_{\Delta=0} 
  & = & \frac{3}{16\pi^2}\frac{\partial \Delta J_1(m,\Delta)}
     {\partial \Delta}\Biggr\vert_{\Delta=0} 
  \to -\frac{3}{16\pi^2}\ell(m^2) \label{eq:hq_int1}\\* 
  (g_{\mu\nu} - v_\mu v_\nu)\frac{\partial}{\partial m^2}
  \frac{\partial J^{\mu\nu}(m,\Delta)}
  {\partial \Delta}\Biggr\vert_{\Delta=0} & = & 
  \frac{3}{16\pi^2}  \frac{\partial}{\partial m^2}\frac{\partial J_1(m,\Delta)}
     {\partial \Delta}\Biggr\vert_{\Delta=0}  
  \to \frac{3}{16\pi^2}\tilde{\ell}(m^2) \label{eq:hq_int2}
\end{eqnarray}
As in Ref.~\cite{SCHPT},
we use the arrow to denote the fact that we drop the analytic
terms and 
keep only the chiral logarithms.
The functions $\ell$ and $\tilde \ell$ are defined as
\begin{eqnarray}\label{eq:chiral_log_infinitev}
        \ell(m^2) &\equiv & m^2 \ln \frac{m^2}{\Lambda_\chi^2}
        \qquad{\rm [infinite\ volume]} \ , \\*
        \label{eq:chiral_log2_infinitev}
        \tilde \ell(m^2)& \equiv& 
        - \frac{\partial}{\partial m^2}\ell(m^2) = -\ln\left(
\frac{m^2}{\Lambda_\chi^2}\right) - 1
\qquad
        {\rm [infinite\ volume]} \ .
\end{eqnarray}
These are the infinite volume forms; 
the effect of finite spatial volume will be discussed in Sec.~\ref{sec:fin_vol}.

For \eq{fB-cur} we need the Minkowski version of the
integrals defined in Ref.~\cite{SCHPT}:
\begin{eqnarray}
  I_1 (m^2) & \equiv & i\Lambda_\chi^{\hat\epsilon} \int \frac{d^{4-\hat\epsilon} p}
{(2\pi)^{4-\hat\epsilon}}\; \frac{1 
    }{(p^2-m^2+i\epsilon)}  \label{eq:lq_int1}
  \to \frac{1}{16\pi^2}\ell(m^2)\\* 
  I_2 (m^2) & \equiv  &i\Lambda_\chi^{\hat\epsilon}  \int \frac{d^{4-\hat\epsilon}
    p}{(2\pi)^{4-\hat\epsilon}} \;\frac{1}{(p^2-m^2+i\epsilon )^2} 
  \to -\frac{1}{16\pi^2}\tilde{\ell}(m^2)\ . \label{eq:lq_int2}
\end{eqnarray}

\section{Final NLO results}
\label{sec:final_results}

Now we are ready to write down the full NLO result, 
including the analytic
terms. 

After performing the integrals using \eqsthru{hq_int1}{lq_int2}, 
we get for the
\opopo\ partially quenched case with all masses unequal:
\begin{eqnarray}\label{eq:1p1p1_pq_fB}
  \left(\frac{f_{B_x}}{f_{B_x}^{\rm LO}}\right)_{\opopo} 
  &= & 1 + \frac{1}{16\pi^2f^2}
  \frac{1+3g_\pi^2}{2}
  \Biggl\{-\frac{1}{16}\sum_{f,\Xi} \ell(m_{xf,\Xi}^2)  \nonumber \\*&&{}-
    \frac{1}{3}\sum_{j\in \cM_I^{(3,x)}} 
    \frac{\partial}{\partial m^2_{X,I}}\left[ R^{[3,3]}_{j}(
      \cM_I^{(3,x)};  \mu^{(3)}_I)\ell(m_{j}^2) \right]
    \nonumber \\*&&{} 
     -  \biggl( a^2\delta'_V \sum_{j\in \cM_V^{(4,x)}}
     \frac{\partial}{\partial m^2_{X,V}}
     \left[ R^{[4,3]}_{j}( \cM_V^{(4,x)}; \mu^{(3)}_V )
    \ell(m_{j}^2) \right]
        + [V\to A] \biggr)
   \Biggr\}      \nonumber \\*&&{}+
        c_s (m_u + m_d + m_s) + c_v m_x + c_a a^2 
\end{eqnarray}
where $f$ runs over the three sea quarks $u$, $d$, and $s$, 
$\Xi$ runs over the 16 meson tastes, 
and the sets of masses in the residues are defined as follows
(with taste labels implicit):
\begin{eqnarray}
  \mu^{(3)} & = & \{m^2_U,m^2_D,m^2_S\}\ ,\\*
  \cM^{(3,x)} & = & \{m_X^2,m_{\pi^0}^2, m_{\eta}^2\}\ ,\\*
  \cM^{(4,x)} & = & \{m_X^2,m_{\pi^0}^2, m_{\eta}^2, m_{\eta'}^2\}\ .
\end{eqnarray}
The analytic terms in \eq{1p1p1_pq_fB} 
come from wave function renormalization contributions 
from $\cL_3$ [\eqsthree{L3m}{L3A}{L3B}],  as well as from
higher order corrections to the current $j_2^{\mu,b}$  [\eqsthree{j2m}{j2A}{j2B}].
The low energy constants (LECs) $c_s$, $c_v$ and $c_a$  can be written as linear combinations
of the large number of constants appearing in $\cL_3$ and $j_2^{\mu,b}$.  However, such expressions are
unlikely to be useful, and we omit them here.

Of relevance to MILC simulations is the case where the up and down
quark masses are degenerate, the $2\!+\!1$ case. Setting $m_u=m_d=m_l$,
we obtain
\begin{eqnarray}\label{eq:2p1_pq_fB}
  \left(\frac{f_{B_x}}{f_{B_x}^{\rm LO}} \right)_{2\!+\!1}
  & = & 1 + \frac{1}{16\pi^2f^2}
  \frac{1+3g_\pi^2}{2}
  \Biggl\{-\frac{1}{16}\sum_{f,\Xi} \ell(m_{xf,\Xi}^2)
  \nonumber \\*&&{}- \frac{1}{3}
\sum_{j\in \cM_I^{(2,x)}} 
\frac{\partial}{\partial m^2_{X,I}}\left[ 
      R^{[2,2]}_{j}(
    \cM_I^{(2,x)};  \mu^{(2)}_I) \ell(m_{j}^2) \right]
    \nonumber \\*&&{} 
     -   \biggl( a^2\delta'_V \sum_{j\in \hat\cM_V^{(3,x)}}
     \frac{\partial}{\partial m^2_{X,V}}\left[ 
       R^{[3,2]}_{j}( \hat\cM_V^{(3,x)}; \mu^{(2)}_V)
    \ell(m_{j}^2)\right]
        + [V\to A]\biggr)  
   \Biggr\} \nonumber \\*&&{}+
        c_s (2m_l + m_s) + c_v m_x + c_a a^2 \ ,
\end{eqnarray}
where 
\begin{eqnarray}
  \mu^{(2)} & = & \{m^2_U,m^2_S\}\ ,\\*
  \cM^{(2,x)} & = & \{m_X^2, m_{\eta}^2\}\ ,\\*
  \hat\cM^{(3,x)} & = & \{m_X^2, m_{\eta}^2, m_{\eta'}^2\}\ .
\end{eqnarray}

Another interesting limit is the case of $N_f$ degenerate flavors of
sea quarks of mass $m_f$. 
\begin{eqnarray}\label{eq:degNf_pq_fB}
  \left(\frac{f_{B_x}}{f_{B_x}^{\rm LO}}\right)_{N_f\;{\rm degen}} 
  &= & 1 + \frac{1}{16\pi^2f^2}
  \frac{1+3g_\pi^2}{2}
  \Biggl\{-\frac{N_f}{16}\sum_{\Xi} \ell(m_{xf,\Xi}^2)  \nonumber \\*&&{}
- \frac{1}{N_f}
    \left[\left(m_{X,I}^2-m_{ff,I}^2\right)\tilde\ell(m_{X,I}^2)  -\ell(m_{X,I}^2) \right]
    \nonumber \\*&&{} 
     -  \biggl( a^2\delta'_V 
     \frac{\partial}{\partial m^2_{X,V}}
     \left[ \frac{\left(m^2_{ff,V}-m^2_{X,V}\right)\ell(m^2_{X,V})
     -\left(m^2_{ff,V}-m^2_{\eta',V}\right)\ell(m^2_{\eta',V})}
{m^2_{\eta',V}-m^2_{X,V}}\right]  \nonumber \\*
&& \hspace{1truecm}+ [V\to A] \biggr)
   \Biggr\}      
        +N_f c_s m_f + c_v m_x + c_a a^2 \ ,
\end{eqnarray}
where, in this case, 
\begin{equation}
m^2_{\eta',V} = m^2_{ff,V} +\frac{N_f}{4} a^2 \delta'_V\ ,
\end{equation}
and similarly for $m^2_{\eta',A}$.
The result in \eq{degNf_pq_fB} reduces to that of Sharpe and Zhang \cite{SHARPE_ZHANG}
in the continuum limit.\footnote{Note that the analytic term coming from
the $-1$ in our \eq{chiral_log2_infinitev} is absorbed into the 
LECs in Eq.~(3.6) of Ref.~\cite{SHARPE_ZHANG}.}
The $1/N_f$ dependence of the taste-singlet term  
in \eq{degNf_pq_fB}  is characteristic of the
disconnected contribution and arises from the fact that 
$m_{\eta',I}\sim N_f\, m_0^2/3$ for large $m_0^2$ in the
$1\!+\!1\!+\!1\!+\!\cdots$ theory.

We now turn to full QCD, where the
valence quark is one of the light sea quarks. In the $2\!+\!1$
case, we have
\begin{eqnarray}
  \left(\frac{f_{{B_u}}}{f_{{B_u}}^{\rm LO}}\right)_{2\!+\!1\;{\rm full}} 
  & = & 1 + \frac{1}{16\pi^2f^2}
  \left(\frac{1+3g_\pi^2}{2}\right)
  \Biggl\{-\frac{1}{16}\sum_{\Xi} \left[ 2\ell(m_{\pi,\Xi}^2) 
    + \ell(m_{K,\Xi}^2) \right]\nonumber\\*&&{}+
    \frac{1}{2}\ell(m^2_{\pi_I}) - \frac{1}{6}\ell(m^2_{\eta_I})  
    \nonumber \\*&&{}
     -   \biggl(a^2\delta'_V \biggl[ 
     \frac{(m^2_{\pi,V} - m^2_{S,V})}{(m^2_{\pi,V} - 
       m^2_{\eta,V})(m^2_{\pi,V} - m^2_{\eta',V})}
     \ell(m_{\pi,V}^2)\nonumber\\*&&{}
     + \frac{(m^2_{\eta,V} - m^2_{S,V})}{(m^2_{\eta,V} -
       m^2_{\pi,V})(m^2_{\eta,V} - m^2_{\eta',V})}
     \ell(m_{\eta,V}^2) \nonumber\\*&&{}+ 
     \frac{(m^2_{\eta',V} - m^2_{S,V})}{(m^2_{\eta',V} 
       - m^2_{\pi,V})(m^2_{\eta',V} - m^2_{\eta,V})}
     \ell(m_{\eta',V}^2)\biggr]
        + [V\to A]\biggr) \Biggr\}  \nonumber \\*&&{}+
        c_s (2m_l + m_s) + c_v m_l + c_a a^2 
   \ , \label{eq:2p1_full_fBu}   \\
  \left(\frac{f_{B_s}}{f_{B_s}^{\rm LO}}\right)_{2\!+\!1\;{\rm full}} 
  & = & 1 + \frac{1}{16\pi^2f^2}
  \left(\frac{1+3g_\pi^2}{2}\right)
  \Biggl\{-\frac{1}{16}\sum_{\Xi} \left[ \ell(m_{S,\Xi}^2)
    + 2\ell(m_{K,\Xi}^2) \right]\nonumber\\*&&{}+
    \ell(m^2_{S,I}) - \frac{2}{3}\ell(m^2_{\eta_I})  
    \nonumber \\*&&{}
     -   \biggl( a^2\delta'_V \biggl[
     \frac{(m^2_{S,V} - m^2_{\pi,V})}{(m^2_{S,V} - 
       m^2_{\eta,V})(m^2_{S,V} - m^2_{\eta',V})}
     \ell(m_{S,V}^2)\nonumber\\*&&{}
     + \frac{(m^2_{\eta,V} - m^2_{\pi,V})}{(m^2_{\eta,V} -
       m^2_{S,V})(m^2_{\eta,V} - m^2_{\eta',V})}
     \ell(m_{\eta,V}^2) \nonumber\\*&&{}+ 
     \frac{(m^2_{\eta',V} - m^2_{\pi,V})}{(m^2_{\eta',V} 
       - m^2_{S,V})(m^2_{\eta',V} - m^2_{\eta,V})}
     \ell(m_{\eta',V}^2)\biggr]
        + [V\to A] \biggr) \Biggr\}  \nonumber \\*&&{}+
        c_s (2m_l + m_s) + c_v m_s + c_a a^2 
   \ . \label{eq:2p1_full_fBs}
\end{eqnarray}
We have used the fact that $m_{\eta,I}^2 =
(2m^2_{S,I} + m^2_{\pi,I})/3$ at this order to simplify the taste-singlet 
residues.  Note that, as expected for full QCD, there are no longer any double-pole contributions.
In the continuum limit \eqs{2p1_full_fBu}{2p1_full_fBs}
reduce to known full QCD results \cite{Goity:1992tp,Grinstein-et,SHARPE_ZHANG}.

\section{Chiral-Scale Dependence and Low Energy Constants}\label{sec:scale}
The effect of a change in the chiral scale $\Lambda_\chi$ can be absorbed into
the LECs $c_s$, $c_v$, and $c_a$.  
In fact, the change of most of the one-loop terms is proportional
to taste-split squared meson masses, rather than quark masses {\it per se}, since it is
the meson masses that appear in the non-analytic terms. It has therefore proved
convenient for light \cite{FPI04} and heavy-light \cite{Aubin:2005ar} lattice calculations
to rewrite the analytic terms as functions of meson masses, rather than quark masses,
and, specifically, to use those combinations of meson masses that arise naturally
from a change in scale in the non-analytic terms.   Since squared meson masses
have additive contributions proportional to $a^2$ due to taste violations 
[see \eq{pi-masses-specific}], this procedure automatically introduces some 
$a^2$ terms proportional to $c_v$ and $c_s$, and is equivalent to a redefinition
$c_a \to C_a = c_a - \beta_v c_v - \beta_s c_s$, where $\beta_{v,s}$ are constants.

The redefinition is convenient because the scale dependence of $C_a$ is simplified
and no longer depends on the splittings $\Delta_\Xi$. More importantly, the
procedure has a practical advantage for reducing
lattice errors. From the simulations, $C_a$ seems to be
significantly smaller than $c_a$.  In other words, most of the discretization error from
the light quarks appears to be due to the $a^2$ dependence of the light meson masses in the chiral
loops.  To the extent this is true, \ie that $C_a$ can be neglected in first 
approximation, the $a^2$
dependence is tied to the mass dependence through $c_v$ and $c_s$.  This means that
a determination of the mass dependence at fixed lattice spacing
can be used to estimate the lattice-spacing dependence.

The appropriate redefinitions do not depend on the quark masses, but, like the LECs
themselves, they do depend on $N_f$.  Therefore it is most convenient 
to work with the case of degenerate sea quark masses but arbitrary $N_f$, \eq{degNf_pq_fB}.
Using \eq{pi-masses-specific},  we can write the
change in the chiral log contribution 
$\delta\! f_{B_{x}}$ [\cf \eq{fB}] under
a change $\Lambda_\chi\to \tilde \Lambda_\chi$
in the chiral scale as
\begin{eqnarray}\label{eq:Del-logs}
  \delta\! f_{B_{x}} 
  &\to  & \delta\! f_{B_{x}} + \log(\tilde\Lambda^2/\Lambda^2)
  \frac{(1+3g_\pi^2)}{2}
  \Biggl\{\frac{N^2_f-4}{2N_f}\biggl(2\mu m_x +a^2 \Delta_{\rm val}\biggr)\nonumber \\
  &&\hspace{3truecm} +\frac{N^2_f+2}{2N_f}\biggl(2\mu m_f +a^2 \Delta_{\rm sea}\biggr)
     +a^2(\delta'_V+\delta'_A) 
   \Biggr\}     \ , 
\end{eqnarray}
where
\begin{eqnarray}\label{eq:Delta-val}
\Delta_{\rm val} &\equiv& \frac{N_f^2}{N_f^2-4}\bar\Delta - \frac{4}{N_f^2-4}\Delta_I \\
\label{eq:Delta-sea}
\Delta_{\rm sea} &\equiv& \frac{N_f^2}{N_f^2+2}\bar\Delta + \frac{2}{N_f^2+2}\Delta_I \ ,
\end{eqnarray}
with $\Delta_I$ the taste-singlet splitting and $\bar\Delta$ the average splitting:
\begin{equation}\label{eq:Delta-av}
\bar\Delta = \frac{1}{16}\sum_\Xi \Delta_\Xi=\frac{1}{16}
\left(4\Delta_A +6\Delta_T +4\Delta_V +\Delta_I\right) \ .
\end{equation}

Based on \eq{Del-logs}, it is natural to redefine the LECs by
\begin{equation}\label{eq:LECs-deg}
N_f\, c_s\, m_f + c_v\,m_x + c_a a^2 \to  N_f\, C_s (2\mu m_f+a^2\Delta_{\rm sea}) + 
C_v(2\mu m_x+ a^2\Delta_{\rm val}) + C_a a^2
\end{equation}
for degenerate quarks [\eg in \eq{degNf_pq_fB}], or
\begin{eqnarray}
c_s (m_u+m_d+m_s+\cdots)+ c_vm_x + c_a a^2 &\to&  C_s \left(2\mu\left(m_u+m_d+m_s+\cdots\right) 
+N_f a^2\Delta_{\rm sea}\right) \nonumber \\
&&\hspace{0.5truecm}+ C_v(2\mu m_x+ a^2\Delta_{\rm val}) + C_a a^2\label{eq:LECs}
\end{eqnarray}
in the general case [\eg in \eq{1p1p1_pq_fB}]. With these redefinitions, the change
in the LECs under a chiral scale change is rather simple:
\begin{eqnarray}
C_s(\tilde \Lambda_\chi) &=& C_s(\Lambda_\chi) - \log(\tilde\Lambda^2_\chi/\Lambda^2_\chi)\,
\frac{(1+3g_\pi^2)}{32\pi^2f^2}\,\frac{(N_f^2+2)}{2N_f^2}\nonumber\\
C_v(\tilde \Lambda_\chi) &=& C_v(\Lambda_\chi) - \log(\tilde\Lambda^2_\chi/\Lambda^2_\chi)\,
\frac{(1+3g_\pi^2)}{32\pi^2f^2}\,\frac{(N_f^2-4)}{2N_f}\nonumber\\
C_a(\tilde \Lambda_\chi) &=& C_a(\Lambda_\chi) - \log(\tilde\Lambda^2_\chi/\Lambda^2_\chi)\,
\frac{(1+3g_\pi^2)}{32\pi^2f^2}\left(\delta'_v+\delta'_A\right) \ .
\label{eq:scale}
\end{eqnarray}

We emphasize that the redistribution of the $a^2$ terms in \eqsor{LECs-deg}{LECs} 
is not unique, and it is possible to move them from $C_s$ to $C_v$ without complicating
the transformation properties of $C_a$ under chiral scale change. In the
``mixed'' meson terms $m_{xf,\Xi}$ in \eq{degNf_pq_fB}, for example, we have chosen 
to associate half of each splitting with the valence quark $x$ and half with the sea
quark $f$. This is simple and ``natural,'' but not required. The freedom stems
from the fact that we have only one condition coming from chiral scale invariance, but
two LECs ($C_v$ and $C_s$) with which we want to associate $a^2$ dependence.
This lack of uniqueness is particularly apparent in the case $N_f=2$, where the
choice we made for $\Delta_{\rm val}$ in \eq{Delta-val} blows up.  The problem
is that, for $N_f=2$, the change in the chiral logs under change in chiral scale has no term
that is proportional to $m_x$ [see \eq{Del-logs}].  In this case, one might wish
to associate all the splittings with the sea quark. Of course, a redefinition like
\eqsor{LECs-deg}{LECs} is completely consistent  for {\it any}\/ choice of
$\Delta_{\rm sea}$ and $\Delta_{\rm val}$.  Some redefinitions are ``better'' than
others, however, if they result in a smaller $C_a$, and our choice does seem to work
in the $N_f=3$ case \cite{Aubin:2005ar}.

\section{Finite Volume Corrections}\label{sec:fin_vol}

In computing the finite volume corrections, we assume that the lattice size in the time direction is
larger than in the spatial directions, and indeed is large enough that it 
may be treated as infinite to a very good approximation.  This is the case
for the MILC lattices \cite{MILC_SPECTRUM,FPI04}.  We are thus interested in
corrections due to finite {\it spatial}\/ volume only.  Furthermore, we assume that the
heavy-light meson is at rest: $v^\mu = (1,0,0,0)$. This is the case in all
lattice simulations of the decay constant except those of ``moving NRQCD'' \cite{MOVING}.

 For integrals of the
the form of \eqs{lq_int1}{lq_int2}, which come from light-meson tadpole loops,
the corrections are expressed in a convenient form in Ref.~\cite{CHIRAL_FSB}. 
The results of these integrals appear in our final answers, \eqsthru{1p1p1_pq_fB}{2p1_full_fBs},
in the terms multiplied by the ``1'' in the factor $1+3g_\pi^2$.
For such terms, we merely need to make the replacements \cite{CHIRAL_FSB}
\begin{eqnarray}\label{eq:chiral_log1_1}
  \ell( m^2)& \to & \ell_1( m^2) \equiv  \ell( m^2)   
  + m^2\delta_1(mL) \qquad{\rm [finite\ volume,\ ``1\ terms"]} \ ,
  \\*
  \label{eq:chiral_log2_1}
  \tilde \ell(m^2)& \to & \tilde \ell_1( m^2) \equiv  
  \tilde \ell(m^2) + \delta_3(mL) 
  \qquad{\rm [finite\ volume,\ ``1\ terms"]} \ ,
\end{eqnarray}
in \eqs{chiral_log_infinitev}{chiral_log2_infinitev}, 
where
\begin{eqnarray}\label{eq:delta1}
        \delta_1(mL) & = & \frac{4}{mL}
                \sum_{\vec r\ne 0}
                \frac{K_1(|\vec r\,|mL)}{|\vec r\,|} \ , \\*
        \label{eq:delta3}
        \delta_3(mL) & =& 2 \sum_{\vec r\ne 0}
                K_0(|\vec r\,|mL)\ ,
\end{eqnarray}
with $K_0$ and $K_1$ the Bessel functions of imaginary argument.

The integrals that involve a heavy-light meson in the loop appear in our
final answers in the terms multiplied by $g^2_\pi$.  For these integrals,
\eqs{hq_int1}{hq_int2}, the finite volume corrections can be determined
by comparison with the corrections to 
\eqs{lq_int1}{lq_int2}.
To do this, we first perform the integrals 
over $p^0$ in both cases.  This integration is not affected by finite volume corrections
since the time extent of the lattices is taken to be infinite.
We obtain
\begin{eqnarray}
  I_1 (m^2) & = & \int \frac{d^3 p}{(2\pi)^3} \frac{1
    }{2\sqrt{\vec{p}\,^2+m^2}}\ ,  \nonumber\\ 
  (g_{\mu\nu} - v_\mu v_\nu)\frac{\partial J^{\mu\nu}(m,\Delta)}
  {\partial \Delta}\Biggr\vert_{\Delta=0} 
  & = & -\int \frac{d^3 p}{(2\pi)^3} \frac{   
    \vec{p}\,^2 }{2(\vec{p}\,^2+m^2)^{3/2}}\nonumber\\*
  & = & -\left[I_1 + 2m^2\frac{\partial I_1}{\partial m^2}  \right]\label{eq:fv-hq}\ .
\end{eqnarray}
Since the finite volume correction to $I_1$ is given by \eq{chiral_log1_1},
the correction to the heavy-light integral \eq{hq_int1} is now determined
\via\ \eq{fv-hq}.
Corrections to the integrals with double poles, \eqs{lq_int2}{hq_int2},
follow by differentiation with respect to $m^2$.

The prescription for finite volume corrections to the terms 
proportional to $g_\pi^2$, is then
\begin{eqnarray}\label{eq:chiral_log1_g}
        \ell( m^2)& \!\to\! & \ell_g( m^2) \equiv \ell(m^2) + \frac{m^2}{3}\left(\delta_1(mL) -
        2\delta_3(mL)\right) \ {\rm [finite\ volume,\ ``}g_\pi^2{\rm\ terms"]}
        \\*
        \label{eq:chiral_log2_g}
        \tilde \ell(m^2)& \!\to\! &  \tilde \ell_g( m^2)\equiv \tilde \ell(m^2) + \delta_3(mL) 
        -\frac{2}{3}\delta_5(mL)
        \quad{\rm [finite\ volume,\ ``}g_\pi^2{\rm\ terms"]},
\end{eqnarray}
where
\begin{equation}
\delta_5(mL) \equiv -\frac{mL}{2}\delta_3'(mL) =  mL \sum_{\vec r\ne 0}
                |\vec r\,| K_1(|\vec r\,|mL)\ , 
\end{equation}
with the prime denoting differentiation with respect to the argument.  We have used 
$K'_0(z)=-K_1(z)$ and the
relation $\delta_3(mL)=-\delta_1(mL)-(mL/2)\delta'_1(mL)$ \cite{CHIRAL_FSB}.

We note that our procedure for finding finite volume corrections
of heavy-light integrals works well only for the rather simple
integrals needed here. It does not easily generalize
to cases where the time extent is finite
or where the variable $\Delta$ in \eq{Jmunu}
is nonzero.
The latter case is particular important since it is needed for semileptonic
form factors, as well as for $1/m_Q$ and $\cO(m_q^2)$ corrections.
The finite size effects for heavy-light \chpt\ have been studied in much more
generality in Ref.~\cite{LIN_ARNDT}.

\section{Remarks and Conclusions}\label{sec:conc}

We have determined the heavy-light \schpt\ Lagrangian to NLO and used it to compute
the heavy-light decay constant in partially quenched and full QCD.
The heavy-light part of the chiral Lagrangian is
identical at leading order to that in the continuum, aside from the extra taste degrees of freedom,
which enter trivially.  Thus taste-violating effects appear in the one-loop chiral
logarithms only through the light-light meson sector, where the relevant parameters
have already been determined by simulations \cite{FPI04,LAT02-03POSTERS,MILC_SPECTRUM}. 
All the operators that appear at NLO in the heavy-light Lagrangian 
serve to produce only a term proportional to $a^2$ in the decay constants, and thus
a single new low energy constant:  $c_a$ [\eqsthru{1p1p1_pq_fB}{2p1_full_fBs}] 
or $C_a$ [\eqsor{LECs-deg}{LECs}].

We emphasize that our calculations are based on the assumption that
the staggered ``fourth root trick'' is valid.  We further assume 
that this trick is correctly implemented on the chiral
theory by insertions of factors of $1/4$ for each sea quark loop, where the
loops are found by quark flow or replica analysis.

Our results most relevant at present are those for the 
$2\!+\!1$ case  ($N_f\!=\!3$ and $m_u\!=\!m_d$), since they
apply to calculations
using existing MILC configurations for the light dynamical quarks. 
These results have proved quite useful in reducing the errors from the
chiral and continuum extrapolations in a lattice calculation of
the leptonic decay constants of $D$ and $D_s$ mesons \cite{Aubin:2005ar}.
With the Lagrangian and the techniques presented here, one may also
find expressions for the form factors for semileptonic heavy-to-light decays (such
as $D\to K\ell\nu$ or $B\to \pi \ell\nu$) in \schpt.
This has been done \cite{Aubin:2004xd}; a detailed report of the calculation
is in preparation \cite{CA-CB-SEMILEPTONIC}.
The result has already been used in Ref.~\cite{Aubin:2004ej}. We also call
attention to the work of Laiho \cite{Laiho:2005np}, who has
applied heavy-light \schpt\ to study heavy-to-heavy semileptonic
decays, \eg $B\to D^*\ell\nu$.

A calculation of heavy-light $B$ parameters in \schpt\ should also be straightforward.
Although additional operators will contribute to analytic terms at
NLO, the one-loop chiral logs themselves will not involve any new 
low energy constants 
beyond those already present in the continuum or in the one-loop \schpt\ for light-light
mesons.  

\begin{figure}
 \includegraphics[width=5in]{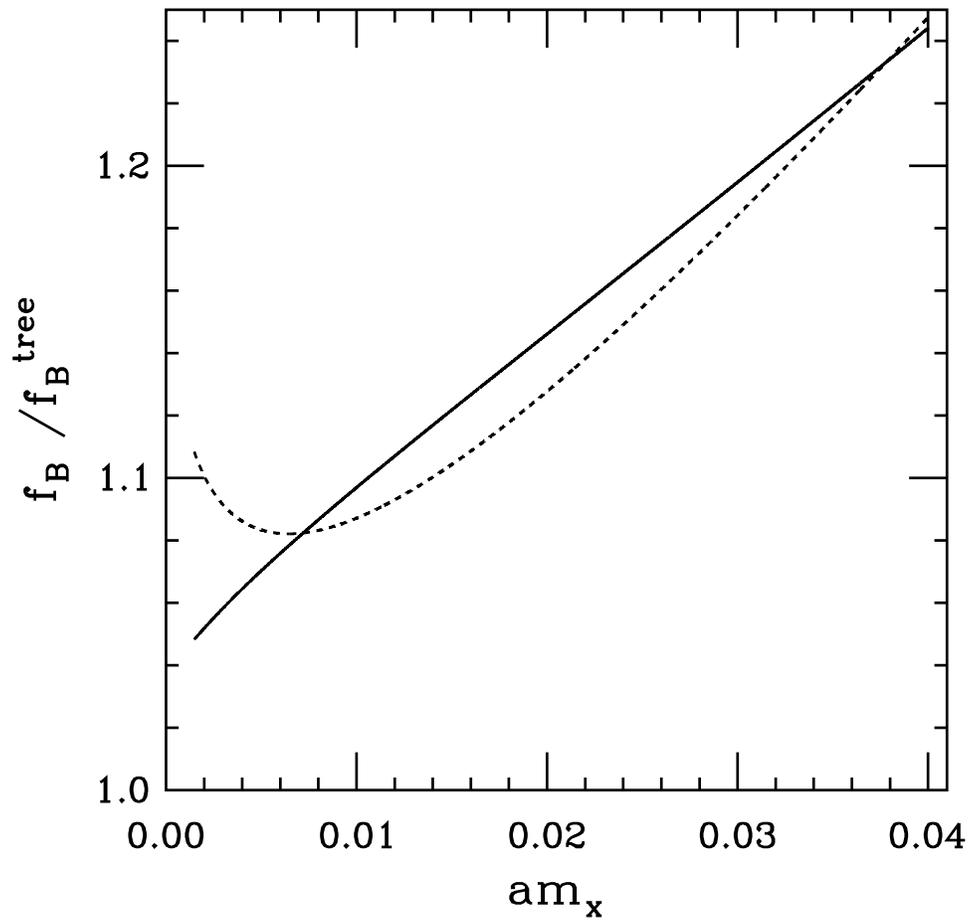} \caption{The ratio of the
 NLO decay constant for two degenerate sea quarks, given by
  \eq{degNf_pq_fB} with $N_f=2$.
The solid line is with the taste violations included,
 while the dashed line is the continuum limit.}
  	\label{fig:fB_am010}
\end{figure}

It is interesting to look at a simple example to see the effect of  taste
violations. In 
Fig.~\ref{fig:fB_am010} we plot $f_{B_x}/f_{B_x}^{\rm LO}$ for the
partially quenched case with two degenerate  sea quarks:  \eq{degNf_pq_fB}
with $N_f=2$. The solid line includes taste-breaking terms;
while the dashed line shows the results with taste-breaking terms
set to zero, \ie the continuum limit. We choose parameters
from the coarse MILC data set ($a\approx
0.125$fm). The light meson masses and splittings,  as well as the values of
$\delta'_{V}$ and $\delta'_A$, come from measurements of
the light hadron spectrum and decay constants 
\cite{FPI04,LAT02-03POSTERS,MILC_SPECTRUM}. We have
used the value $am_f = 0.010$ for the sea quark
mass. We set $c_s=c_a=0$ and choose $c_v$ to 
give a slope at large mass similar to what is seen in simulations.
The continuum result shows the characteristic divergence of partially quenched
chiral perturbation theory as the valence mass is taken to zero for fixed sea
mass.  The taste violations, however, completely wash out this behavior and give
a result that looks rather linear.  This is in fact an example 
of an infrared sensitive quantity for
which the chiral and continuum limits in \schpt\ will not
commute \cite{Bernard:2004ab}.
Fitting to and removing the effects of staggered taste violations is therefore
crucial in controlling 
the systematic error associated with the extrapolation to the physical values of
the light quark masses.

\bigskip
\bigskip
\centerline{\bf ACKNOWLEDGMENTS}
\bigskip
We thank J.\ Bailey, B.\ Grinstein, A.\ Kronfeld, P.\ Mackenzie, S.\ Sharpe and 
our colleagues in the MILC collaboration for 
helpful discussions. This work was partially supported by the
U.S. Department of Energy under grant number DE-FG02-91ER40628.

\vfill\eject

\end{document}